\documentclass[useAMS,usenatbib]{mn2e}
\usepackage{graphicx}
\usepackage{subfigure}
\usepackage{longtable}
\usepackage{lscape}

\usepackage{color}

\definecolor{Magenta}{cmyk}{0.1,0.8,0,0.1} 
\definecolor{Green}{rgb}{0.0, 0.5, 0.0}
\newcommand{\bhc}{4U~1630-472\  }
\newcommand{\bhcc}{4U~1630-472.\  }
\newcommand{\igr}{IGR~J16320-4751\ }

\title[The puzzling case of the BHC 4U~1630-472]{Missing hard states and regular outbursts: the puzzling case of the  black hole candidate \bhc}

\author[F. Capitanio et al.]{F. Capitanio$^{1}$\thanks{E-mail:
fiamma.capitanio@iaps.inaf.it}, R. Campana$^{2}$, G. De Cesare$^{2}$, C. Ferrigno$^{3}$\\
       $^{1}$ INAF-IAPS, via Fosso del Cavaliere, 100, 00133, Roma, Italy\\
       $^{2}$ INAF-IASF, Via Piero Gobetti, 101, 40129 Bologna, Italy\\
       $^{3}$  ISDC Data Centre for Astrophysics, Chemin d'Ecogia 16, 1290 Versoix, Switzerland\\ 
}
\begin{document}

\date{}

\pagerange{\pageref{firstpage}--\pageref{lastpage}} \pubyear{2013}

\maketitle

\label{firstpage}

\begin{abstract}

\bhc  is a recurrent X-ray transient classified as a black-hole candidate from its spectral and timing properties. One of the peculiarities of this source is the presence of regular outbursts with {a recurrence period between  600 and 730 d that has been observed  since the discovery of the source in 1969.} We report on a comparative study on the spectral and timing behaviour of three consecutive outbursts occurred in 2006, 2008 and 2010, respectively. We analysed all the data collected by the  {\it INTErnational Gamma-Ray Astrophysics Laboratory} (INTEGRAL) and the {\it Rossi X-ray timing Explorer }(RXTE) during these three years of activity. We show that, {in spite of having a similar spectral} and timing behaviour in the energy range between 3 and 30 keV,  these three outbursts show pronounced differences above 30 keV. 
In fact, the 2010 outburst extends at high energies without any detectable cut-off until 150--200 keV, while the  two previous outbursts occurred in 2006 and 2008 are not detected at all above 30 keV.  Thus,  in spite of a very similar accretion disk evolution, these three outbursts exhibit totally different characteristics of the Compton electron corona, showing a softening in their evolution rarely observed before in a low mass X-ray binary hosting a black hole.   We argue the possibility that the unknown perturbation that causes the  outbursts to be equally spaced in time could be at the origin of this particular behaviour. {Finally we describe several possible scenarios that could explain the regularity of the outbursts, identifying  the most plausible,  such as a third body orbiting around the binary system.}

\end{abstract}

\section{Introduction}
\label{se0}
\bhc is one of the most active black hole candidates (BHCs) ever observed. The first known outburst was detected by Vela\,5B  satellite in 1969 \citep{Priedhorsky}. Since then source has undergone {an} outburst at least 20 times up to now and has been observed by most of the principal X-ray missions which have collected a huge
amount of data.

\bhc is a highly absorbed source with a hydrogen column density {that varies in the range of $N_H=(4-12)\times10^{22}$ cm$^{-2}$~\citep[][]{Tomsick,Kuulkers}}.  
The distance and the mass of the compact object have still not been firmly measured because no optical counterpart has been identified up to now.
This is mostly due to the crowded region in which the source lies. Thus,  \bhc is still classified as a BHC {because of its characteristic spectral and timing behaviour.

\citet{Seifina} have recently reported an indirect estimation of the mass (10 M$_{\odot}$) on the basis of the basis of the detection of the spectral index saturation with the mass accretion rate~\cite[see also][]{Shaposhnikov}. }
{The infrared (IR) counterpart  was instead detected during the 1998 outburst as reported by~\citet{Augusteijn}. The analysis of the IR {photometric} data reveals that \bhc is a highly reddened source lying in the direction of a giant molecular cloud that is located at a distance of 11 kpc. \bhc lies in the near side of this cloud thus at a distance $<$ 11 kpc. However the high absorption indicates a large distance, $>$ 10 kpc~\citep{Seifina}.}  {Thus, it is reasonable to assume a source distance of $\sim$10--11 kpc.}
The IR properties are consistent with  a relatively long orbital period system (P$_{orb}$~$\sim$ few days) containing an early type secondary star. Anyway, the range in absolute K-band magnitude {and the observed IR variability} cannot exclude a  B star nature (intrinsically slightly reddened) in a detached Be/X-ray binary
~\citep{Augusteijn}.

 {\citet{Dieters} and ~\citet{Trudolyubov} report on detailed X-ray spectral and timing analysis of the  1998 outburst of \bhcc They show that the source behaviour is typical of a black hole candidate. During the same outburst, the detection of a polarized radio emission revealed the presence of optically thin radio jets ejected from the source~\citep{Hjellming}. 
  
  Several radio observation campaigns have been carried on during the subsequent outbursts.  Nevertheless}, most of them failed to detect any radio emission from the source~\citep{Hannikainen,Gallo,Calvelo}.

\citet{Tomsick} report the detection of a dip in the {\it Rossi X-ray Timing Explorer (RXTE) Proportional Counter Array (PCA)} data during the 1996 observations. This implies that the inclination angle is probably greater than 60$^\circ$. 
{\it Suzaku} X-ray satellite observations in 2006 revealed significant absorption lines from highly ionized iron lines with a blueshift that corresponds to an outflow velocity of about 1000 km~s$^{-1}$~\citep{Kubota}.  This {fact supports the hypothesis that the source is viewed at a high inclination angle~\citep{Ponti},
even though, after 1996, no more dips have been reported in literature. {The scaling technique~\citep[see][and references therein]{Shaposhnikov} permitted~\citet{Seifina} to constrain the inclination angle of the system at $i\leq70^\circ$}~\citep{Seifina}. 

{\citet{Diaz Trigo} report on the detection of Doppler-shifted X-ray emission lines  in coincidence with the
reappearance of radio emission from the jets during the 2012 outburst. 
They argue that these lines arise from baryonic matter in a relativistic jet. However, \citet{Neilsen} show that } the radio emission during the same outburst,  was not always associated with relativistic emission lines, and the unique behavior reported by \citet{Diaz Trigo} might be a special case, dependent on additional processes in the accretion flow around the Black Hole (BH). 
\citet{King} report a clear detection of reflection component in a short  observation performed with the NuSTAR telescope during the 2013 outburst when the source was in an intermediate state. The consequent spin measurement indicates that \bhc harbors a rapidly spinning BH.

A unique characteristic of \bhc is that the outbursts
that have been observed from the first detection in 1969 have
been roughly equally spaced in time,
 with a period of about 600 d~\citep{Parmar}.  As the \bhc long-term RXTE/{PCA} light curve shows (Figure~\ref{HR_tot}, top panel), the recurrence is still present after more than 40 yr, {even though a drift in the recurrence period of about 130 days has been observed in the last 3  outbursts, 2008-2012~\citep{Capitanio_1}.  A similar drift in the recurrence period between one outburst and a subsequent outburst was previously noticed by~\citet{Kuulkers1}}.
In addition, \bhc is known to show peculiar outbursts that often lack bright hard states~\citep[see e.g.][and reference therein]{Tomsick14, Abe}. 

The aim of this paper is to perform a comparative study of the high-energy behaviour of 3 consecutive outbursts of \bhc that occurred from 2006 until 2010. In Section~\ref{se2} we introduce the data collection criteria,  the data reduction methods and the techniques employed for spectral and timing analysis; in Section~\ref{se3} we describe the general behavior of the source during the three outbursts. In Section~\ref{se7} we discuss the implication of the peculiar behaviour of the three outbursts in the light of the standard {theory} of the accretion flow on to a low-mass X-ray binary (LMXRB) hosting a BH. In the same section we describe the possible scenarios that could also explain the periodic recurrence of the outbursts. Finally,  we summarize our finding and conclusions in Section~\ref{se8}.

\section{Data Reduction and Analysis}
\label{se2}
The 2008--2010  outbursts of \bhc were extensively observed by RXTE and INTEGRAL satellites. {More than 800 ks of observations have been collected by the RXTE PCA, while 850 ks have been collected by  the INTEGRAL $\gamma$--ray telescope IBIS~\citep{Ubertini}}.

{The RXTE/PCA observation campaign  covered the period of the three outbursts from MJD 53727 to MJD 55459 for a total of 441 pointings.}
The PCA analysis was performed with the standard RXTE software within {\sc HEASOFT} 12.7.1 following   standard extraction procedure for light curves and spectra.
For the PCA spectral analysis a systematic error of 0.6 per cent~\citep{Wilms} was added to the spectra. For the fitting procedure, the energy range 3--30 keV was used.

The timing analysis of the PCA was performed with custom software: for each of the observations, we produced power spectra from 16-s stretches accumulated in the channel band 0--35 (2--15 keV) with a time resolution  of 1/125 s. 
The resulting power spectra were then averaged in one power spectrum per observation, normalized according 
to~\citet{Leahy} and then converted into squared fractional rms ~\citep{Belloni90,Myamoto}. The contribution due to Poissonian statistics was subtracted as reported by~\citet{Zhang}.

For the INTEGRAL data analysis, we used the latest release of the standard Offline Scientific Analysis, OSA version 10.0, distributed by the INTEGRAL Science Data Centre ~\citep[ISDC,][]{Courvoisier} . 
The INTEGRAL analysis was focused on the low-energy detector~\citep[ISGRI,][]{Lebrun} of the $\gamma$-ray telescope IBIS. 
The ISGRI spectra were extracted in 20--200 keV energy range. A systematic error of 2 per cent  was taken into account for spectral analysis~\citep[see also][]{Jourdain}. In order to collect enough counts in a single spectrum we added together contiguous IBIS  spectra showing the same shape.  {We obtained 13 IBIS spectra: each of which was  quasi-simultaneous with a single PCA observation. }
{The faintness of the source  in most of the data and the limited energy range  (3--30 keV)  
  allowed us to modelize  the 441 PCA spectra of the three outbursts only with a simple model, that is, an absorbed multicolor disk black body~\citep{Sakura} plus a power law to model the Comptonization of soft photons in a hot plasma: {\sc wabs$\times$(diskbb+pow)}, hereafter {\sc model 1}. 
  We also attempted to model the  Comptonization, instead of power law, with  two different simple physical models:   {\sc nthcomp}~\citep[][]{Zdziarski96}  or {\sc compTT}~\citep{Titarchuk}, {\sc wabs*(diskbb+nthcomp)} and {\sc wabs*(diskbb+comptt)}, hereafter {\sc models 2} and {\sc 3}, respectively. { Because the cut--off is not constrained by the data, we do not  succeed in constraining the plasma temperature. Thus we fixed the electron plasma temperature at 15 keV for 2006 and 2008 outburst spectra  and at 100 keV for 2010 spectra, respectively (see Sections~\ref{se4},~\ref{se5} and~\ref{se6} for details)}. The temperature of the seed photons was, instead, fixed at the same value of the multicolor disk black body model in the case of the {\sc model 2}, while for {\sc model 3,}  {  the seed photon temperature was fixed for each observation  at its best fit value}. The statistics, the energy range and the resolution of the spectra did not allow us to apply any other more refined models. }
\begin{figure*}
\includegraphics[scale=0.7,angle=90]{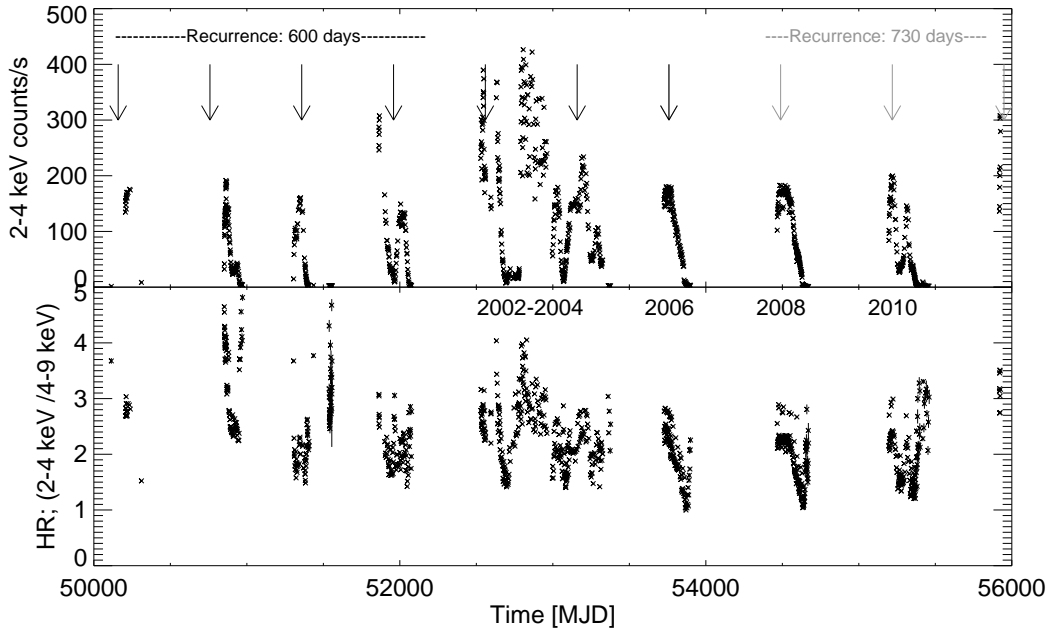}
\caption{{Top panel:  2--4 keV} PCA long term light curve of \bhcc Each point corresponds to a single PCA observation. Bottom panel: HR {(4--9 keV/2--4 keV)} versus time as observed by RXTE/{ PCA} during all its observational life (1996-2011). The black arrows show the 600 d recurrence period of the outbursts, while the grey arrows show the 730 d one (see Section~\ref{se0} for details).}
\label{HR_tot}
\end{figure*}

 {Because of the limited PCA low energy coverage, the equivalent hydrogen column, {\it N$_{H}$}, was allowed to vary between a range of values extracted from the different results reported in  the literature~\citep[see e. g.][]{Tomsick05,Trudolyubov}, (4--12)$\times$10$^{22}$~cm$^{-2}$. 

In order to improve the fit, we added a Gaussian component to the model in some of the observations, especially the ones where the source is particularly faint. This is not only due to an intrinsic Fe emission line but also due to the Galactic ridge emission line contribution not completely corrected in the PCA data and to a nearby source, \igr, which has a prominent iron emission line (see Section~\ref{se21} for details). {The $\chi^{2}$ statistic was used to determine the goodness of the fit}. Only 4 per cent of the PCA pointings were  immediately excluded because of an unacceptable $\chi^{2}$/degree of freedom (d.o.f.) 
( 10 per cent  in the case of the Comptonization models ).  The excluded data mostly belongs to the last part of the three outbursts when the source is very faint and the contamination due to the nearby source can no longer be neglected (see Section~\ref{se21} for details) 
 } 
 {Thus the accepted data have a $\bar{\chi}_{red}^{2}$=1.0 with a maximum spread of 0.5$<$$\chi_{red}^{2}$ $<$1.9} and {with the 76 per cent of the values that lie in the range: 0.8$<$$\chi_{red}^{2}$ $<$1.2. Figure~\ref{chi2} shows the distribution of the ${\chi}_{red}^{2}$ values of the accepted PCA data. All the errors and upper limits reported in the paper correspond to the 90 per cent confidence level. 
 
 \begin{figure*}%
\includegraphics[scale=0.5,angle=0]{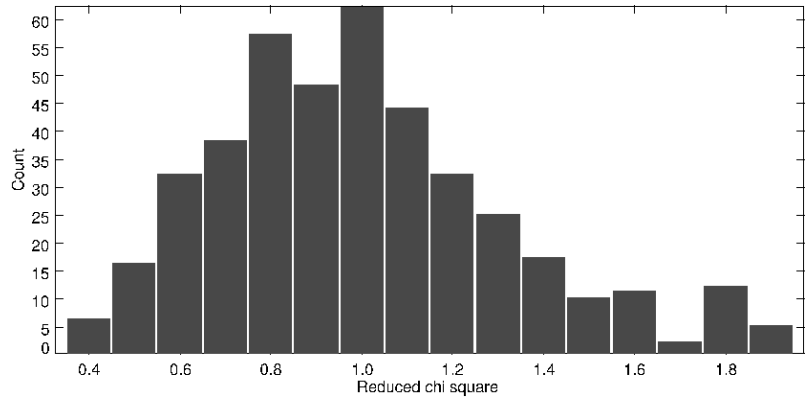}
\caption{Distribution of the reduced $\chi^{2}$ values of the accepted PCA data fit.}
\label{chi2}
\end{figure*}

{A normalization constant was added in the data model used to fit the 13 PCA--IBIS joint spectra to consider the calibration of the
two different instruments. Moreover, because of their broad-band (3--200 keV), they were also modeled taking into account more refined models.}

First, we applied a more refined Comptonization model such as {\sc compps}~\citep{Poutanen} in order to model the high energy part of the joint PCA-IBIS spectra. However any attempt to  discriminate between thermal, non thermal or hybrid Comptonization, did not succeed  in constraining the parameters (see Section~\ref{se6} for details). 

 Then we considered the possible presence of reflection component~\citep{Ross}. The model used was an absorbed multicolor disk blackbody~\citep{Sakura} plus a reflected power law in neutral medium~\citep[{\sc pexrav}]{Magdziarz} that leads to a rapid computation~\citep{Fabian}: {\sc wabs*(diskbb+pexrav)}, hereafter {\sc model 4}.  Because the PCA-IBIS joint spectra do not show any cut-off the {\sc pexrav} parameter linked to the energy cut-off was fixed to zero as required by the model.}

\subsection{Contamination of the RXTE data by nearby source, \igr.}
\label{se21}
The PCA collimated field of view (FOV) has a radius of 1$^\circ$. 
Thus, as also reported by ~\citet{Tomsick14}, the PCA data of \bhc could be contaminated by the nearby persistent X-ray source \igr, which is a well-studied High Mass X-Ray  Binary (HMXRB) discovered by INTEGRAL in 2003~\citep[see e.g.][]{Rodriguez} and lies 0$^\circ$.25 away from \bhcc

During the 2006 outburst, the {\it Joint European X-Ray Monitor} (JEM-X) of INTEGRAL  observed the \bhc  field for a total of 14 ks. \bhc was detected in the 2--10~keV mosaic image at flux level of 270 mCrab; while \igr was not detected at all (upper limit {of a few} mCrab, {see left panel of Figure~\ref{image}}). 
	{The latter source was instead clearly detected in the IBIS mosaic image of the same INTEGRAL observations at flux level of 14 mCrab, while \bhc was} not detected at all {(see Figure~\ref{image}, right panel)}.
Looking at  the All-Sky Monitor (ASM) and the Burst Alert Telescope (BAT) light curves of \igr we can see that, even though it is variable, the source is faint and does not show any  particular increase of the flux during the three \bhc outbursts.
\begin{figure*}
\includegraphics[scale=0.65,angle=0]{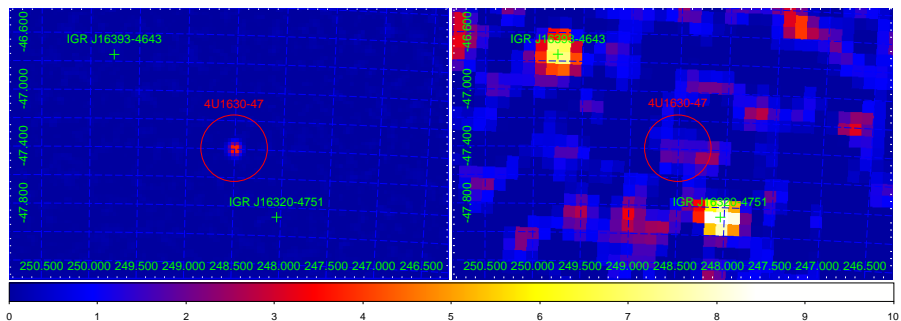}
\caption{ left panel: 3-7 keV 2006 outburst JEM-X mosaic image of \bhc and \igr sky field (exposure 14 ks   for a total of 6 science windows that belong to revolutions 399, 407, 411, 419, 423; MJD interval: 53756--53826). Right panel: 15-30 2006 outburst IBIS mosaic image of \bhc and \igr sky field (exposure 31 ks  for a total of 14 science windows that span from revolution 399 until revolution 427; MJD interval: {  53756--53840)} }
\label{image}
\end{figure*}

We extracted the predicted PCA count/s using the \texttt{ppms} tool and the detailed spectral analysis reported by \citet{Rodriguez}. We considered the {less favorable} case in which the source is in flaring state with a flux of 2$\times$10$^{-10}$erg~s$^{-1}$cm$^{-2}$~\citep{Rodriguez}. 
Considering also the PCA response curve as a function of the off axis position~\citep{Jahoda}, we obtained a PCA count rate of 10 counts~s$^{-1}$, a rate consistent with \citet{Tomsick14}.
{ Thus, when \bhc approaches 10 PCA count/s the contribution of \igr  is no longer negligible}, but it  can be considered almost constant. {In fact the flux variation of \igr is quite limited and  the source varies in flux but not in the spectral shape~\citep{Rodriguez}}. {The source spectra also present a gaussian emission line, centered at 6.4 keV, and an iron edge~\citep{Rodriguez}.}  
Considering the continuum spectral shape of \igr flaring state as reported by~\citet{Rodriguez}.  and the continuum spectral shape of \bhc during the last PCA observations, 
we put a limit on spectral reliability {at a} conservative flux level of 4$\times$10$^{-10}$ erg~s$^{-1}$cm$^{-2}$ in the 2--10 keV energy range (which correspond to $\sim$50 counts~s$^{-1}$). Figure~\ref{sottraspe} shows the comparison between \igr flaring spectrum as reported by~\citet{Rodriguez} and the 50 counts/s \bhc spectrum  of the 2006 outburst.

During both 2006 and 2008 outbursts, \bhc was not detected above 30 keV by INTEGRAL/IBIS { and Swift/BAT (see, for example, the BAT light curve in Figure~\ref{asm_bat} and right panel of Figure~\ref{image}  for the 2006 IBIS outburst image.}). Thus the RXTE/High Energy X-ray Timing Experiment (HEXTE) spectra are irreparably contaminated by the \igr which instead, shows a bright emission {above} 30 keV and therefore these have not been used in the subsequent analysis. During the 2010 outburst RXTE/HEXTE was no longer in operation.

\begin{figure}%
\includegraphics[scale=0.6,angle=0]{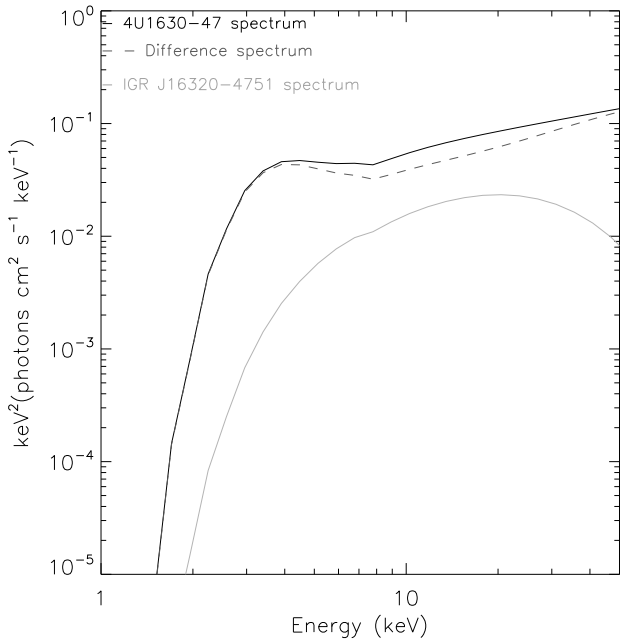}
\caption{ Comparison between the 2010 outburst unfolded spectral model of \bhc  at flux level of 4$\times$10$^{-10}$ erg~s$^{-1}$cm$^{-2}$ (that corresponds to $\sim$50 counts/s in the 2--10 keV energy range) and the flaring state of  \igr  (from~\citet{Rodriguez}). The plot also shows the difference between the two spectra (dark-grey dashed line).
}
\label{sottraspe}
\end{figure}

\section{Results}
\label{se3}
 As Figure~\ref{asm_bat} shows, the source light curve in the energy range 2--20 keV is very similar for the first two outbursts in 2006 and 2008, while the 2010 outburst presents a different time evolution showing two bright peaks in the light curves and reaching harder energy values.  Figure~\ref{HID} shows the RXTE/PCA Hardness-Intensity diagrams (2--20 keV versus 9--20 keV/2--9 keV, hereafter HID) of the three outbursts, while Figure~\ref{curvaluce} shows the PCA light curve in two energy ranges (first two panels), the Hardness Ratio (HR) evolution as a function of time (third panel) and  the  evolution of the root mean square (\textit{rms}) variability as a function of time (last panel). The dashed lines represent the time when the source goes below a PCA source flux of 50 counts~s$^{-1}$: after this  conservative limit the contamination of \igr can no longer be ignored.
   As Figures~\ref{HID} shows, the HID of the 2006 and 2008 outbursts are mostly identical, while the HID of the 2010 outburst evolves at higher energies. The 2010 outburst reaches  a 30 per cent grater values with respect to those reached during the 2006 and 2008 outbursts. 

\begin{figure*}
\includegraphics[scale=0.5,angle=90]{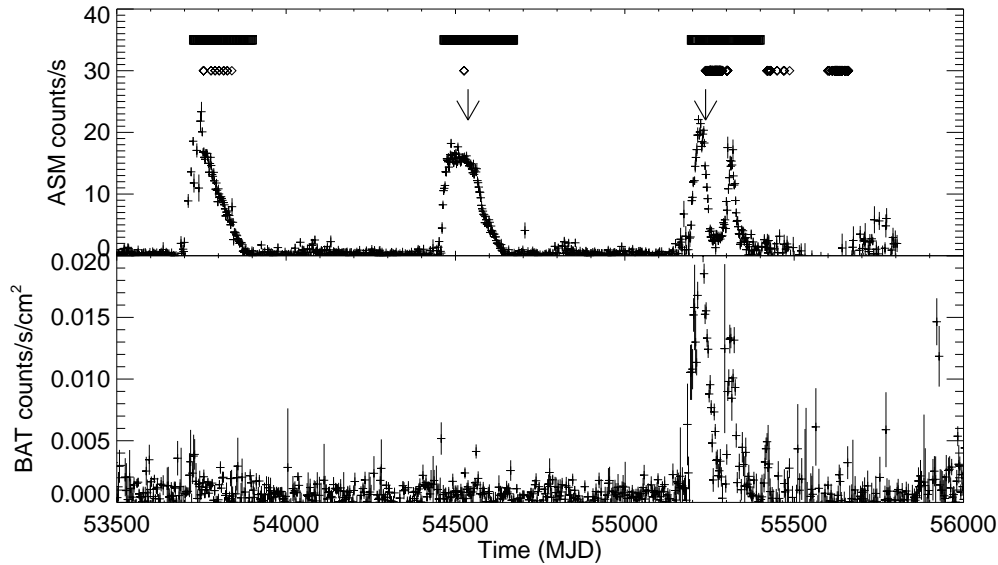}
\caption{ Top panel: \bhc ASM light curve (bin size 2 d). The rectangles represent the PCA observation dates. The diamond points represent the INTEGRAL observations; the arrows correspond to the spectra in Figures~\ref{spettri1}--\ref{spettri22} (see Section~\ref{se3.4}). Bottom panel: 15--50 keV {\it Swift}/BAT light curve (bin size 3 d). \bhc BAT 1$\sigma$ detection sensitivity is 5.3 mCrab for a full-day observation, 1 mCrab is 0.00022 count~cm$^{-2}$s$^{-1}$~\citep{Krimm}.}
\label{asm_bat}
\end{figure*}

\begin{figure*}
\includegraphics[scale=0.6,angle=0]{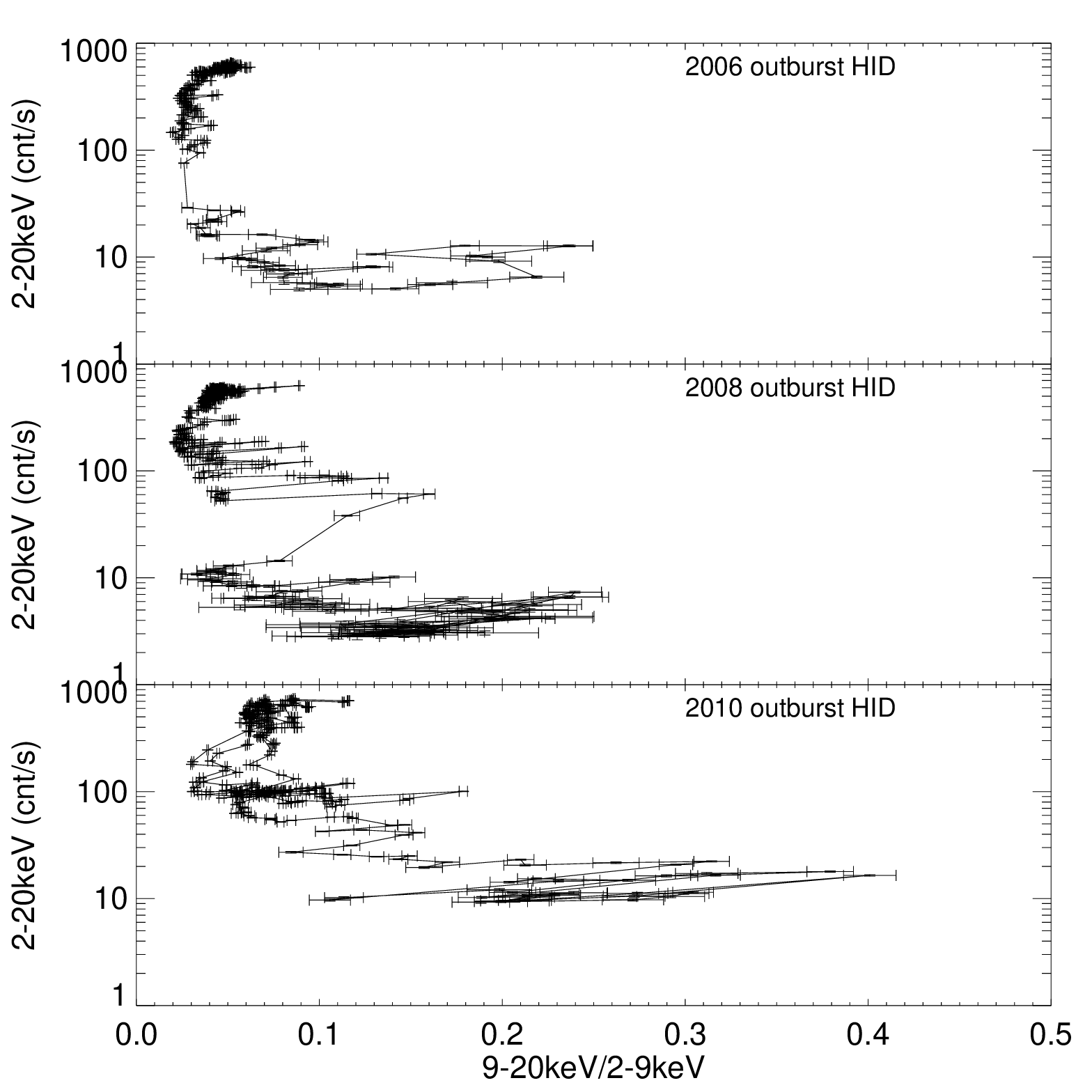}
\caption{Hardness Intensity Diagrams for the three outbursts: 2006 (top panel), 2008 (middle panel) and 2010 (bottom panel). }
\label{HID}
\end{figure*}

 No hard X-ray emission is significantly detected for the 2006 and 2008 outbursts as shown by the light curve taken from the BAT hard X-ray transient monitor on board the {\it Swift} satellite~\citep{Gehrels} (Figure~\ref{asm_bat}). In contrast, the BAT light curve of the 2010 outburst shows a flux about 14 times higher than that of the two previous outbursts.

\begin{figure*}
\includegraphics[scale=0.7,angle=0]{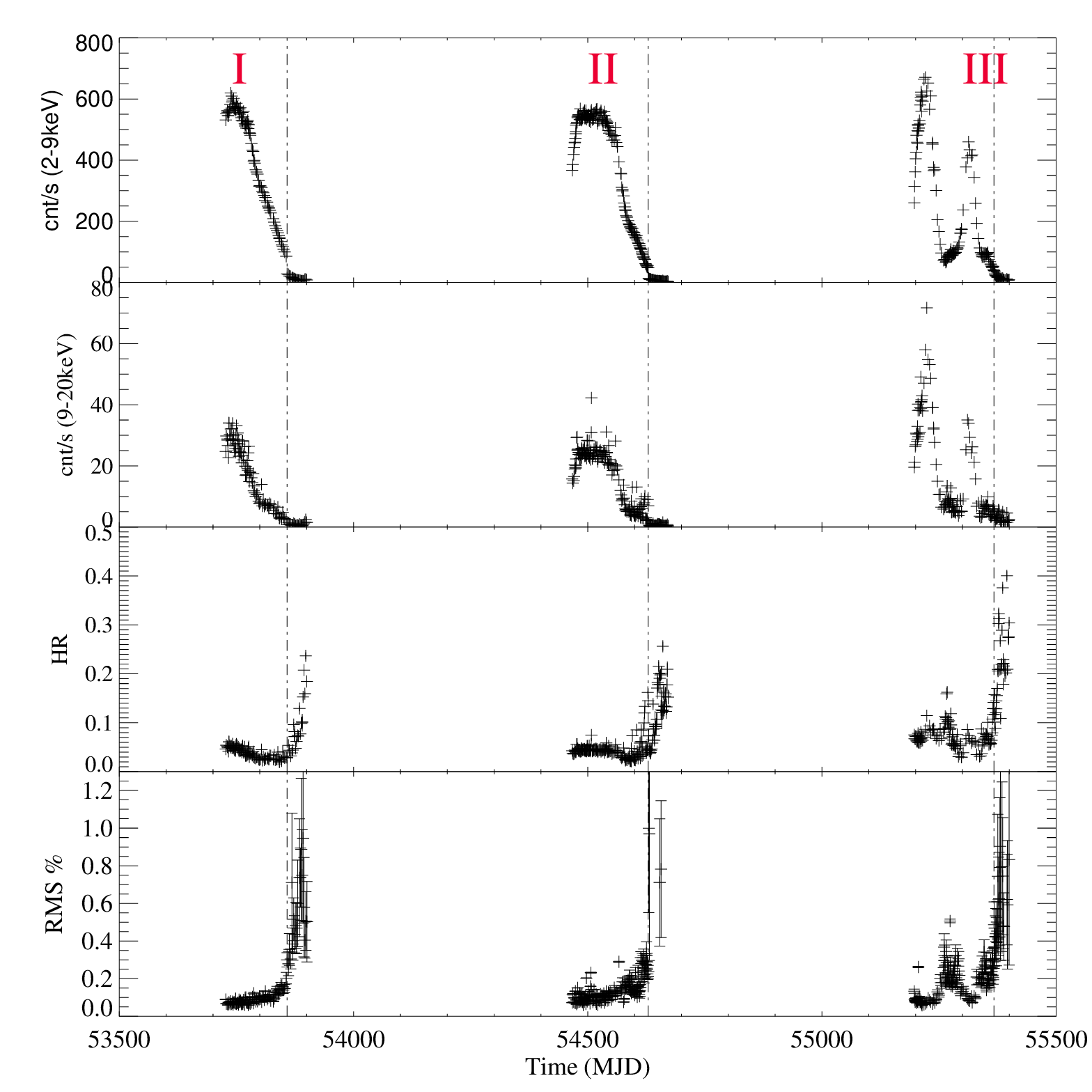}
\caption{ PCA light curve of the 3 outbursts (I,II,III). Top panel: 2--9 keV; upper middle panel 9--20 keV; bottom middle panel: HR (9--20 keV/2--9 keV) versus time; bottom panel: fractional rms versus time. Each point corresponds to a PCA pointing. The dashed lines represent the time when the source reaches a count rate of 50 counts/s in the 2--10keV energy range. After this limit the contamination due to \igr cannot be ignored anymore.}
\label{curvaluce}
\end{figure*}

The INTEGRAL data give similar results: no detections are present in the mosaic image of {the} IBIS data for both 2006 and 2008 outbursts. The IBIS upper limit is 3 mCrab in 20--40 keV (for a total exposure time of 31 ks) for the 2006 outburst and 5 mCrab for the 2008 outburst (25 ks), respectively. The diamond points in Figure~\ref{asm_bat} (top panel) show the INTEGRAL coverage of the 3 outbursts. In contrast, the JEM-X 2006 mosaic data show \bhc as a moderately bright source, 270 mCrab in 3--7 keV energy range. The 2008 {outburst is outside the} JEM-X field of view. However the ASM monitor give consistent results (the peak luminosity is at about 300 mCrab in 1--12 keV).

The IBIS mosaic image of the 2010 outburst shows \bhc as a moderately bright source at flux level of 20 mCrab (20--40 keV) and 83 mCrab (40--100 keV). {The 2010 outburst is also outside the JEM-X FOV.}

   {Considering the ~\citet{Fender} classification - see also ~\citet{Belloni_1} for more details and~\citet{Remillard} for an alternative classification of spectral states - the shapes of the power spectra,  extracted from the PCA data  taken during the three outbursts, are dominated by a power law and show that the source remains substantially in the high soft  (HSS) or intermediate soft state (see Figures~\ref{spettri1}-\ref{spettri22} and Tables~\ref{table2} and~\ref{table3}). However,  because of the faintness of the source, power spectra resolution does not allow us to determine the details of the  power density spectrum (PDS)}.

\subsection{ Evolution of the 2006 outburst}
\label{se4}

The RXTE 2006 observation campaign started on 2005 December 23 (MJD 53727), about 15 d after the ASM flux enhancement~\citep{Tomsick05_1}.
Simultaneous observations with the Australia Telescope Compact Array (ATCA) performed on {  2005} December 21, 25, 27 and 29,  do not detect any radio counterpart of \bhc on any of the four dates~\citep{Gallo}.
There is no evidence of enhanced flux in the BAT light curve all over the outburst (Figure~\ref{asm_bat})  and no detection in the IBIS data.
 Thus {the spectral} analysis was focused on the PCA data collected during the 2006 outburst observation campaign.

After a brief enhancement, the  2--9 keV and 9--20 keV PCA fluxes decrease monotonically during the outburst as shown on the first two panels of Figure~\ref{curvaluce}.
 {The dashed line in Figure~\ref{curvaluce} represents the limit of 50 counts~s$^{-1}$}. As we have seen in Section~\ref{se21}, after this limit the contamination of the nearby source \igr cannot be ignored. This corresponds to the last 23 pointings. The contamination due to \igr is the cause of the huge hardening of the data in Figure~\ref{curvaluce} (bottom middle panel). In any case, a slight hardening of the data (before the dashed lines) can be attributed to \bhcc Nevertheless, even in the last observations, when the power law photon index decreases in value and the source spectra are harder, {the F-test strongly indicates that a disk black body component is still needed in the spectral models {to provide a good fit }(F-test probabilities $<$10$^{-5}$). }
 
 The evolution of the most significant fit parameters along the 2006 outburst is plotted in Figure~\ref{spe_tot_new}(a). {The disk black body emission dominates the spectrum for two and half months (T$_{in}\sim$1.3 keV). Then,} the inner temperature of the disk starts to decrease together with the flux  { reaching inner temperatures of about $\sim$0.8 keV. Thereafter, the contamination of the nearby source can no longer be neglected. The power law photon index, instead, reaches extremely soft values even considering the huge errors. This happens when the high energy flux (20--30 keV) is below 4$\times$10$^{-11}$~ergs~cm$^{-2}$s$^{-1}$.  In these spectra the power-law component - even if required in order to obtain a good fit - is so small that a systematic error in the determination of the photon index cannot be excluded.

During the second part of the outburst  the power law photon index} hardens and remains mostly constant at values of $\sim$1.8. Moreover, there is no evidence in the data of a power-law cut-off until 30 keV. {These spectral parameters are compatible with a soft state as indicated by the {  PDS}, 

However, because the PCA spectra reach only 30 keV and there are no detections of the source in IBIS and BAT data (both instrument share a similar upper limit of few mCrab), we can impose, with good confidence, that there should be a cut-off at energies around 30 keV. To demonstrate this fact, Figure~\ref{new} shows, as an example, the PCA spectrum of the observation 91704-03-42-00 plus the IBIS upper limit. This Figure clearly shows that the lack of  a cut-off should imply a detection at least by one of the two instruments, IBIS and BAT.

{We also applied to the high energy part of the data, instead of the power law, two simple physical models describing Comptonization  of soft photons in a hot plasma, {\sc model 2} and {\sc model 3} (see Section~\ref{se2}). Because of the lack of cut-off in the data, the electron plasma temperature parameter could not be defined by the fitting procedure. 

\begin{figure}
\includegraphics[scale=0.4,angle=0]{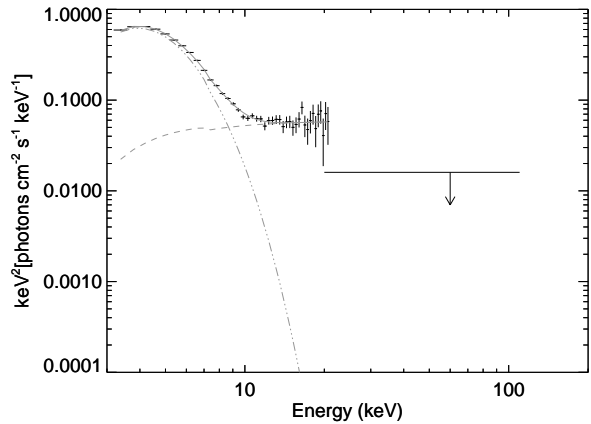}
\caption{{  PCA unfolded spectrum of the 2006 outburst plus the IBIS upper limit (3 mCrab), obs. ID 91704-03-42-00. For model and fitting parameters see Table 2. The dashed lines represent the single model components} }
\label{new}
\end{figure}
{    

Then, because the energy cut off  is roughly  E$_{c}\sim$2--3 kT$_{e}$~\citep[see e.g][and references therein]{Miyakawa,Petrucci}, we fixed the plasma temperature at 15 keV. With this condition the}  spectral parameters of {\sc model 2} are consistent within the errors with those of {\sc model 1}. {  The {\sc model 3}  plasma optical depth increases together with the hardening of the source with values within the range  0.4--4.3 (see Figure~\ref{spe_tot_new}), whereas,} the softest spectra are mostly dominated by the disk black body and the plasma optical depth is quite low ($\sim$ 1). It is worth noticing that the {\sc compTT} model, used to model the high energy part of the spectra in {\sc model 3} is not valid for simultaneously low temperatures and low optical depth. }{  In fact for kT$_{e}$=15 keV and low values of $\tau$ ($\le$ 1) the corresponding values of $\beta$ parameter~\citep{Titarchuk} are outside the zone of applicability of the model ~\citep[][Figure 7]{Hua}. }

%

\begin{figure*}
\includegraphics[scale=0.75,angle=90]{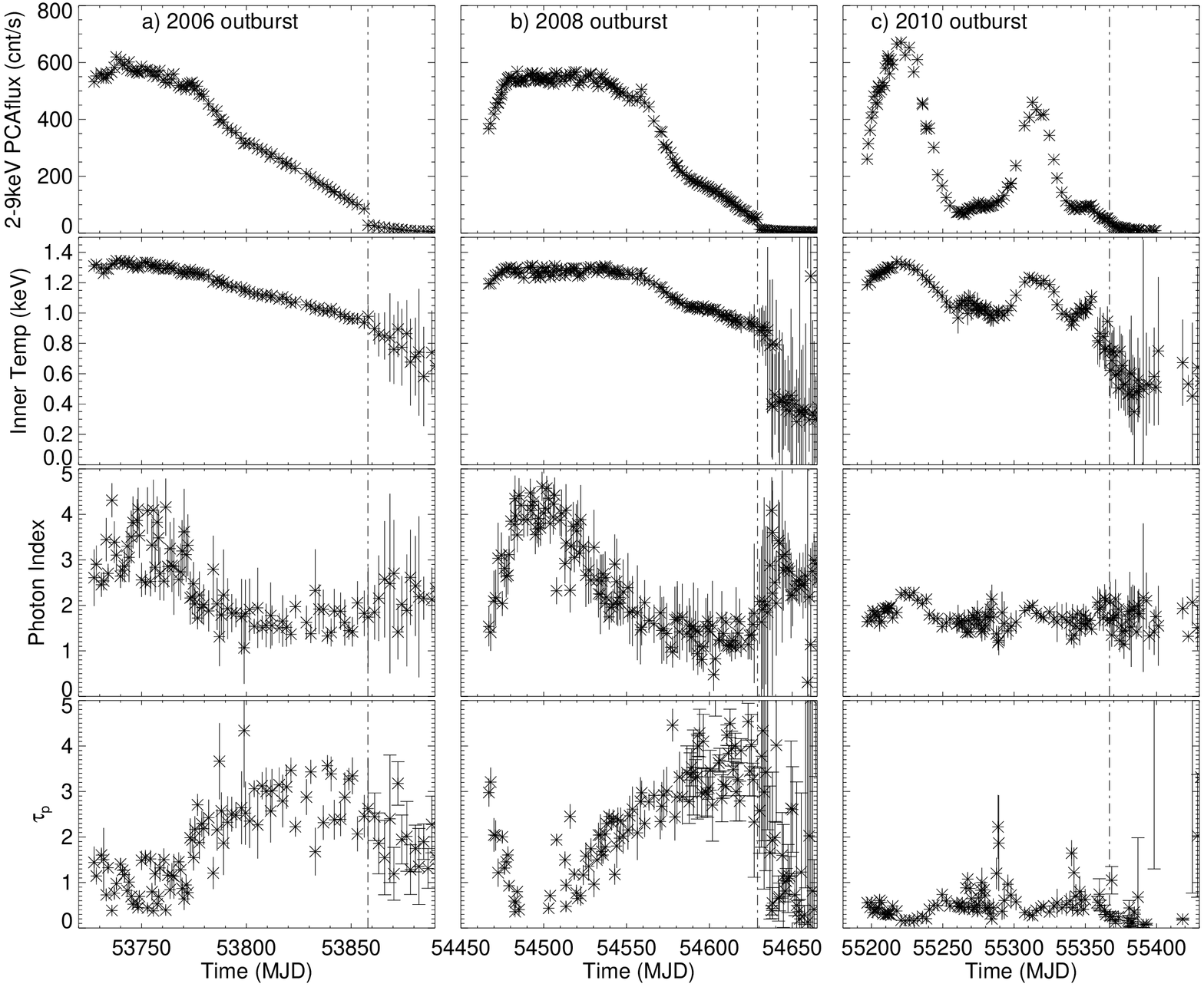}
\caption{  2006, 2008, 2010 outburst flux and fit parameters evolution of {\sc model 1}: {\sc wabs$\times$(diskbb+pow)}. ({ $\bar{\chi}_{red}^{2}$=1.0, 0.5$<$$\chi_{red}^{2}$ $<$1.9}).  {The bottom panels represent the evolution of the Plasma optical depth, $\tau_{p}$ of  {\sc model 3}: {\sc wabs$\times$(diskbb+compTT)}.}
The dashed lines represent the flux limit after which the contamination of the nearby source, \igr, cannot be neglected. Each point of the Figures represents  the value of a spectral parameter of a single PCA observation.}
\label{spe_tot_new}
\end{figure*}

\subsection{ Evolution of the 2008 outburst}
\label{se5}
The RXTE observation campaign started on 2008 January 01 (MJD$\sim$54466), about 6 d {  after the first ASM detection}~\citep{Kalemci}. There is no evidence of both flux enhancement in the BAT light curve  and detection in the IBIS data.
Thus {no hard X-ray emission is detected above 30 keV as in the case of the 2006 outburst. Also for this outburst, the spectral analysis was focused on all the data} collected from PCA during the 2008 outburst.

The evolution of the two most significant fit parameters during the whole 2008 outburst is plotted in Figure~\ref{spe_tot_new} (panel $b$).  The behaviour of this outburst substantially retraces the 2006outburst:  {after a rapid enhancement, the disk black body emission dominates the spectrum for the first three months (T$_{in}\sim$1.3 keV). Then   }the inner temperature of the disk black body  decreases monotonically during all the outburst together with the source flux. {As for the previous outburst,  during the softest part of the spectral evolution, the power law photon index reaches extremely soft values. This happens when the disc black body emission dominates the spectrum and the high energy flux (20--30 keV) is below 3$\times$10$^{-11}$~ergs~cm$^{-2}$s$^{-1}$. } There is no evidence in the data of a power law cut-off until 30 keV, {even though, in analogy with the 2006 outburst, the lack of IBIS and BAT detection indicates a cut-off just above 30 keV} {  (see e. g. Figure~\ref{spettri1}).} 

During the last PCA pointings, there is evidence {  of an increase of both the rms and the HR values} (Figure~\ref{curvaluce}).
However {the F-tests strongly indicates that in} these observations a disk black body component is needed in the spectral models {to obtain a good fit} (F-test probability $<$ 10$^{-3}$). The spectral parameters as reported in Figure~\ref{spe_tot_new} (panel $b$) are compatible with a soft state of the source as indicated by the PDS.
  In the last 45 pointings the contamination of the nearby source cannot be ignored ({ dashed line} in Figure~\ref{spe_tot_new}, panel $b$). 
  
{We also applied to the data the same models used for the 2006 outburst observations ({\sc model 2} and {\sc model 3} as defined in Section~\ref{se2}), fixing the plasma temperature at 15 keV.  As for the previous outburst the {\sc model 2} spectral parameters are consistent within the errors with those of  {\sc model 1}, whereas {\sc model 3} plasma optical depth increases together with the hardening of the source. The softest spectra are mostly dominated by the disk blackbody, thus it is impossible to fit the high energy data even with a simple Comptonization model.  {  As the bottom panel of Figure~\ref{spe_tot_new} shows, during the 2008 outburst the optical depth reaches values comparable with the 2006 outburst (0.4 $< \tau_{p} <$ 4.5). Also in this case values of $\tau$ lower than 1 are outside the zone of applicability of the model~\citep{Hua}.}}

\subsection{ Evolution of the 2010 outburst}
\label{se6}
On 2009 December 29 (MJD 55194), the gas slit camera (GSC) of the Monitor of All-sky X-ray Image (MAXI) detected an increase in X-ray intensity from \bhc ~\citep{Tomida}. Five days later, a radio observation campaign was carried out with the ATCA telescope consisting of two pointings (1440 s per observation). {Once again, also during the 2010 outburst, the radio emission of the source} was below the detection limit of the instrument (93 $\mu$Jy and 126 $\mu$Jy at 5.5 and 9 GHz respectively)~\citep{Calvelo}. 
The RXTE observation campaign started on 2010 January 01. This outburst was  observed by INTEGRAL for a total IBIS exposure time of about 700~ks simultaneous with RXTE. 

The PCA spectral parameters evolution reported in Figure~\ref{spe_tot_new} ($c$) and the enhancement of the emission above 15~keV (see {\it Swift}/BAT light curve in Figure~\ref{asm_bat}) give an indication of two subsequent {softenings of the source with an increase of both the inner temperature and the power law photon index (Figure~\ref{spe_tot_new}, panel $c$)}. {Nevertheless}, during the {source hardenings, subsequent to the two soft peaks, the contribution of the disk emission remains prominent according to~\citet{Tomsick14} and the average photon index value remains at $\sim$1.7}.

 {Just to the end of the 2010 outburst  and in spite of the \igr contamination, 
 the PCA data show  a slight hardening of the source
  (see e. g. HR and the rms in Figure~\ref{asm_bat}). No power law cut-off is detected in the IBIS data up to {  150--200} keV.  Therefore the greatest difference between this outburst and the previous two is the bright emission above 30 keV (Figure~\ref{asm_bat}).}
 
{We also applied to the high energy part of the spectra, instead of a power law, the same comptonization models used for the {  two} previous outbursts. In this case we fixed the plasma temperature at 100 keV because of the BAT continuos detection of the source during 2010 outburst and because of the {  PCA-IBIS joint spectra fit results that does not allow us to constrain any cut-off until 150--200 keV. However we could not exclude that,  when IBIS was not observing the source, a cut-off could be present at energies lower than 200 keV but anyway greater than 30 keV.}
The {\sc model 2} spectral parameters are consistent within the errors with those of {\sc model 1}. Whereas the bottom panel of Figure~\ref{spe_tot_new} shows {\sc model 3} plasma optical depth evolution. The optical depth slightly increases when the spectra hardens but its variation is less important with respect to the values reached during the two previous outbursts.
  
  {  Even though we do not consider the high energy cut-off,} the accuracy of the joint PCA--IBIS spectra is not sufficient to discriminate between thermal, non-thermal or hybrid electron distribution using more refined Comptonization models such as {\sc compps}.

}
Because of the broad-band spectra and the higher emission above 30 keV, {we attempt to fit the joint PCA--IBIS data  {of the 2010 outburst taking into account the reflection component. The model used for the fitting procedure has been introduced in Section~\ref{se2} as {\sc model 4.}}}   The joint spectra (see top panel of Figure~\ref{asm_bat}) are  mostly concentrated during the declining phase of the first soft peak of the {2010} outburst.  Table~\ref{integral_t} reports the evolution of {all} the PCA--IBIS spectral parameters.
 {The PCA--IBIS spectral resolution is not {sufficient to constrain} the reflection component well, which has large errors or can be even defined only with an upper limit. For the spectra where the reflection component is constrained by the data (even if with huge errors) the F-test probability indicates that the fit is improved (F-test probabilities $<$4$\times$10$^{-3}$). {An alternative view reported by~\citet[][]{Laurent} and based on bulk motion photons propagation, demonstrates, using Montecarlo simulations, that for $\Gamma >$ 2  there is no bump in the reflection spectrum due to down-scattering accumulation of photons from high energy tail of the incident spectrum. 

Our results obtained from the 13 joint PCA-IBIS spectra are consistent with $\Gamma \leq$ 2 with the exception of the first observation for which the reflection component is not constrained ({see Table~\ref{integral_t}}).  Even though far from conclusive, our results are supported by~\citet{King} who reports a clear presence of the reflection component in the intermediate state of the 2013 source outburst during a short {  NuSTAR} observation.
 
 }} 

The INTEGRAL observations continued after the end of the RXTE pointings. The last line of Table~\ref{integral_t} shows that the source persisted in having a spectrum extending at high energies that can be fitted with good confidence with a power law ($\Gamma\sim$2) without cut-off until 150--200 keV. Unfortunately, there are no data below 20 keV because the source was outside the JEM--X monitor FOV.

\begin{table*}
\begin{center}
\renewcommand{\thefootnote}{\alph{footnote}}

\caption{{ 2010 outburst spectral evolution of the 13 PCA--IBIS joined spectra of \bhc}. Best fit model: {\sc const*wabs*(diskbb+pexrav)}. Note: {\it Rev} is the IBIS revolution number; {N$_{H}$ is the equivalent hydrogen column (free to vary in the range 4--12$\times$10$^{22}$ cm$^{-2}$, previously reported in literature~\citep[see e. g.][]{Tomsick05,Trudolyubov}); }T$_{in}$ is the temperature of the inner disk; {  $N_\mathrm{disk} = ((R_in/\mathrm{km})/(D/\mathrm{kpc}))^2$} is the normalization parameter of the Sakura\& Sunyaev disk black body; $\Gamma$ is the power law photon index; Rel$_{Refl}$ is the reflection scaling factor {(the cosine of inclination angle is let fixed at its default value, 0.45)}. {F$_{2-10keV}$ is the unabsorbed flux between 2--10 keV. F$_{10-100keV}$ is the unabsorbed flux between 10-100 keV.$\chi$$^{2}$$_{r}$ is the reduced $\chi$$^{2}$. The last line show the spectral parameter of the averaged IBIS spectrum of the last part of the outburst fitted with a simple power law (see Section~\ref{se6})}}
\label{integral_t}
\resizebox*{0.97\textwidth}{!}{\begin{tabular}{lcccccccccc}
\hline
\hline

Rev         & Obs. Date    &  IBIS exp.&        N$_{H}$              &                   T$_{in}$                       &       N$_{disk}$       &  $\Gamma$            &    Rel$_{refl}$           & F$_{2-10keV}$  & F$_{10-100keV}$ & $\chi$$^{2}$$_{r}$(d.o.f.) \\
--&          MJD &         ks   &    10$^{22}$atm~cm$^{-2}$ &   keV                            &        --                         & --                            &   --                       & ergs~cm$^{-2}$s$^{-1}$ & ergs~cm$^{-2}$s$^{-1}$& - \\
\hline
895         & 55239   &  14 &   7.3$_{ -0.3}^{+0.3}$     &      1.22$_{ -0.02}^{  +0.02}$&    190$_{ -21}^{+24}$&  2.2$_{-0.1}^{  +0.1}$&    $<$0.2               & 5.1e-09  & 1.4e-09 &0.8(77) \\
896         & 55241   &  18 &   8.1$_{ -0.3}^{+0.3}$     &      1.20$_{ -0.01}^{  +0.01}$&    177$_{ -18}^{+20}$&  2.1$_{-0.1}^{  +0.1}$&    0.5$_{ -0.3}^{+0.3}$ & 5.1e-09  & 1.4e-09&1.1(66)\\
897         & 55244   &  15 &   7.8$_{ -0.3}^{+0.3}$     &      1.16$_{ -0.01}^{  0.01 }$&    170$_{ -19}^{21}$   &   2.0$_{-0.1}^{  +0.1}$&    0.5$_{ -0.3}^{+0.4}$ & 4.1e-09  & 1.1e-09&1.1(70)\\
897,898  & 55247   &  16 &   6.4 $_{ -0.4}^{ +0.4}$   &      1.12$_{ -0.02}^{  +0.02}$&    126$_{ -19}^{+22}$&  2.0$_{-0.1}^{  +0.1}$&    0.4$_{ -0.3}^{+0.4}$& 2.7e-09  & 8.8e-10 & 1.4(73)\\
898         & 55248   &  14 &   6.3 $_{ -0.3}^{ +0.3}$   &      1.10$_{ -0.02}^{  +0.02}$&    120$_{ -21}^{+24}$&  1.7$_{-0.1}^{  +0.2}$&    $<$0.3               & 2.2e-09  & 7.3e-10& 1.2(73)\\
899,900     & 55252   &44&   6.2 $_{ -0.5}^{ +0.5}$   &      1.04$_{ -0.02}^{  +0.02 }$&   100$_{ -18}^{+23}$&  2.0$_{-0.1}^{  +0.1}$&    0.6$_{-0.3}^{+0.5}$  & 1.5e-09  &5.2e-10 & 1.2(74)\\
901         & 55257   &  26 &   5.3$_{ -0.5}^{+0.5}$     &      1.03$_{ -0.02}^{  +0.02}$&     84$_{ -19}^{+25}$&  1.8$_{-0.1}^{  +0.2}$&    $<$0.4               & 1.1e-09  & 4.0e-10& 1.1(70)\\
901,902     & 55259   &32&   5$_{-1}^{+1}$              &      0.98$_{ -0.04}^{  +0.03}$&     86$_{ -25}^{+36}$&  2.2$_{-0.3}^{  +0.2}$&    1.5$_{-0.9}^{+1.2}$  & 1.0e-09  & 3.7e-10& 1.3(72)\\
902,904,905 & 55265&16&   $<$4             &      1.1$_{   -0.1}^{  +0.1}$&      32$_{ -11}^{+17}$   &  2.0$_{-0.1}^{  +0.1}$&    0.6$_{-0.4}^{+0.5}$  &  6.7e-10 & 4.3e-10&0.8(72)\\
906,907     & 55272   &72&   5.4$_{ -0.6}^{+0.6}$    &      1.03$_{ -0.03}^{  +0.03}$&    105$_{ -23}^{+31}$ &  2.1$_{-0.1}^{  +0.1}$&    0.8$_{-0.4}^{+0.5}$  & 1.6e-09  & 4.9e-10 & 1.0(70)\\
907,908     & 55276   &79&   5.8$_{ -0.5}^{+0.6}$   &      1.01$_{ -0.02}^{  +0.02}$&    111$_{ -22}^{+28}$ &  1.9$_{-0.1}^{  +0.2}$&    0.8$_{-0.4}^{+0.6}$  &  1.3e-09 &3.8e-10& 1.4(70) \\
909,910,911 & 55284&128& 7$_{-1}^{ +1}$            &      0.9$_{ -0.03 }^{  +0.03 }$&   121$_{ -36}^{+47}$ & 2.0$_{-0.1}^{  +0.2}$&    $<$0.6               &  1.3e-09 &2.6e-10& 0.8(72) \\
916         & 55302     &  95&   8.5$_{-0.4}^{+0.4}$   &      1.1 $_{ -0.01}^{  +0.01 }$&   266$_{ -38}^{+45}$ &  1.9$_{-0.1}^{  +0.1}$&    $<$0.4               &4.7e-09& 6.1e-10& 0.8(70)\\
956-959-965& 55420-55449 & 78 & --               &               --               & --                  & 2.1$_{-0.3}^{  +0.3}$      & --                 &    --   &2.3e-10&0.6 (26)\\

\hline
\hline
\end{tabular}
}
\end{center}

\end{table*}

\subsection{Comparison among the 3 outbursts}
\label{se3.4}
 The top panel of Figure~\ref{HR_tot} shows the long term light curve between 2 and 4 keV of all the outbursts detected by the RXTE/PCA from 1996 until 2011, while the bottom panel shows the HR, defined as the ratio between the 4--9 keV and the 2--{ 4} keV energy ranges, versus time.
The bottom panel of Figure~\ref{HR_tot} indicates that the 2006 and 2008 outbursts reached softest HR values with respect to the 2010 outburst, and {also with respect to} all the previous outbursts observed by RXTE including the outbursts with similar luminosity (e.g. the 2001 outburst, MJD$\sim$52000 in Figure~\ref{HR_tot})\footnote[1]{The ASM HR vs time plot gives the same results.} {that are reported in that literature as standard outbursts~\citep[][]{Dieters,Trudolyubov} }

{  In order to highlight the differences between the two 2006 and 2008 outbursts and the 2010 outburst, we present, as an example, the energy  and the power spectrum of two PCA observations; the first was performed during the 2008 outburst} and the second during the 2010 one.  Figure~\ref{spettri1} shows the energy spectrum of the 2008 PCA observation 93425-01-11-07 and the 25~ks IBIS upper limit; Figure~\ref{spettri11} shows the power spectrum of the same PCA observation.
{ Figure~\ref{spettri2} shows, instead, the energy  spectrum of the 2010 PCA} observation 95360-09-24-00 with the simultaneous IBIS spectrum, while Figure~\ref{spettri22} shows the relative {PCA power spectrum.

The two PCA observation dates} correspond to the black arrows in Figure~\ref{asm_bat} and exhibit the same source state: similar spectral shape between 2 and 20 keV, similar {\it rms} value, similar power spectrum and similar 2-20 keV flux emission {as Tables~\ref{table2} and~\ref{table3} show}.   

 Above 30 keV, instead, there is a pronounced difference. In fact, for the 2010 outburst, the data above 30 keV can be fitted with a power law and  no cut--off is required in the model to fit the data up to 250 keV. { Also for the 2008 outburst, a power law component is needed in the model to obtain a good fit} until 30 keV. However, {as we have demonstrated in Sections~\ref{se5} and~\ref{se5} ,} the non-detection of the source at higher energy range implies {instead that  there should be a cut off just above 30 keV. Moreover, the huge difference in value, between the power law normalization parameter of the two spectra,  indicates that } in the case of the 2008 outburst the contribution of the power law in the spectrum is lower than in the case of the 2010 outburst, while the apparent inner radius extracted from the normalization constant of  {\it diskbb} is 26 and 30 km, respectively, see for the correction factor between the apparent inner disk radius and the realistic radius~\citet{Kubota98}.

\begin{figure}
\includegraphics[width=9cm,angle=0]{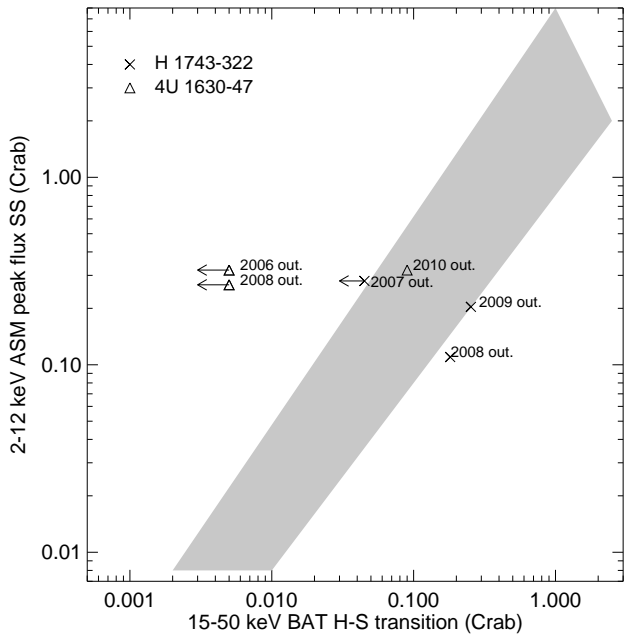}
\caption{BAT flux during the hard to soft transition versus the corresponding ASM flux of 4U1630-47 and H 1743-322 as defined in~\citet{Yu2}. The grey zone represent the region populated by other BHXRB reported by~\citet{Yu2}. } 
\label{plot_wu}
\end{figure}

 {All the three} outbursts seem not to follow the two empirical relations reported by ~\citet{Yu2}, ~\citet{Yu} and ~\citet{Wu} between the hard X-ray  luminosity peak of the HS to HSS transition and the HSS luminosity peak, and between the hard luminosity peak and the outburst waiting time, respectively. 
{ In fact, following the relation reported by~\citet{Wu}, a constant waiting time, which is the peculiarity of \bhc, should imply a constant hard X-ray luminosity peak of the subsequent outburst. This relation is not verified at least between the 2008 and 2010 outbursts, or between the 2002-2004 and 2006 outbursts (Figure~\ref{HR_tot}). Concerning the relation occurring between the X-ray HS to HSS transition luminosity peak versus the subsequent HSS luminosity peak, Figure~\ref{plot_wu} shows that the 2006 and 2008 outbursts are totally outside the correlation reported by~\citet{Wu} (grey filled zone in Figure~\ref{plot_wu}). In fact, the HS--HSS transition should have occurred before the start of the PCA observation campaign. On the contrary, the 2010 outburst is almost in agreement, emphasizing that the lack of hard X-ray emission in the case of the 2006 and 2008 outbursts is unusual.\footnote[2]{ We consider as the Hard to Soft transition flux peak of the 2010 outburst the BAT 15-50 keV luminosity taken on 2009-12-29 that corresponds to the first detection of the outburst~\citep{Tomida}. In fact just two days later on 2009-12-31, the source exhibits a spectral behaviour typical of HSS~\citep{Tomsick09} }

{In order to understand the uniqueness of  \bhc outburst recurrence and to correlate it with the lack of high energy emission, {we investigated} the ASM and BAT light curves of some of the brightest and active transient BHXRBs during the seven years (2005-2011) in which both instruments were operational. Many studies on the outburst evolution of these sources have been reported in literature (e. g. ~\cite{Yu2,Dunn,Capitanio_1}, and references therein); the various outbursts observed do not have any detectable recurrence period and in addition, show a complex behaviour in both soft and hard energy ranges and very different luminosities.  
  The only exception is H 1743-322: this BHC has shown only for some years outbursts equally spaced in time, as reported by~\citet{Capitanio_1}. In analogy with \bhc, the outburst of H 1743-322 that occurred in 2007, has a 2-12 keV behaviour mostly identical to  some of the subsequent  periodical outbursts but showing a fainter hard X-ray emission:  
  the H 1743-322 ASM 2007 peak flux  (1-12 keV) is $\sim$250 mCrab; 
  the BAT 2007 peak flux (15-50 keV) is $\sim$60 mCrab.  The subsequent H 1743-322 {  outburst (2008-2010) presents} a similar 1-12 keV flux but with a 15-50 keV flux of $\sim$ 200 mCrab. 
As Figure~\ref{plot_wu} shows, the difference in the ratio between BAT HS--HSS transition luminosity and ASM peak flux of H 1743-322 is not enough to bring the outburst totally out of the correlation reported by~\citet{Yu2} as in the case of the 2006 and 2008 outbursts of \bhcc However, as reported by ~\citet{Corbel} the HS--HSS transition occurred before the peak of the hard X-ray emission and thus  it is only an upper limit.     For the 2007 and the 2008-2010 outbursts of H 1743-322, also the relation between the hard X-ray luminosity peak and the outburst waiting time is not respected. In contrast, during the subsequent outbursts, a similar waiting time corresponds to a similar peak luminosity in the 15-50 keV energy range.}}


\begin{figure*}
 \begin{minipage}[b]{8cm}
 \includegraphics[width=9cm,angle=0]{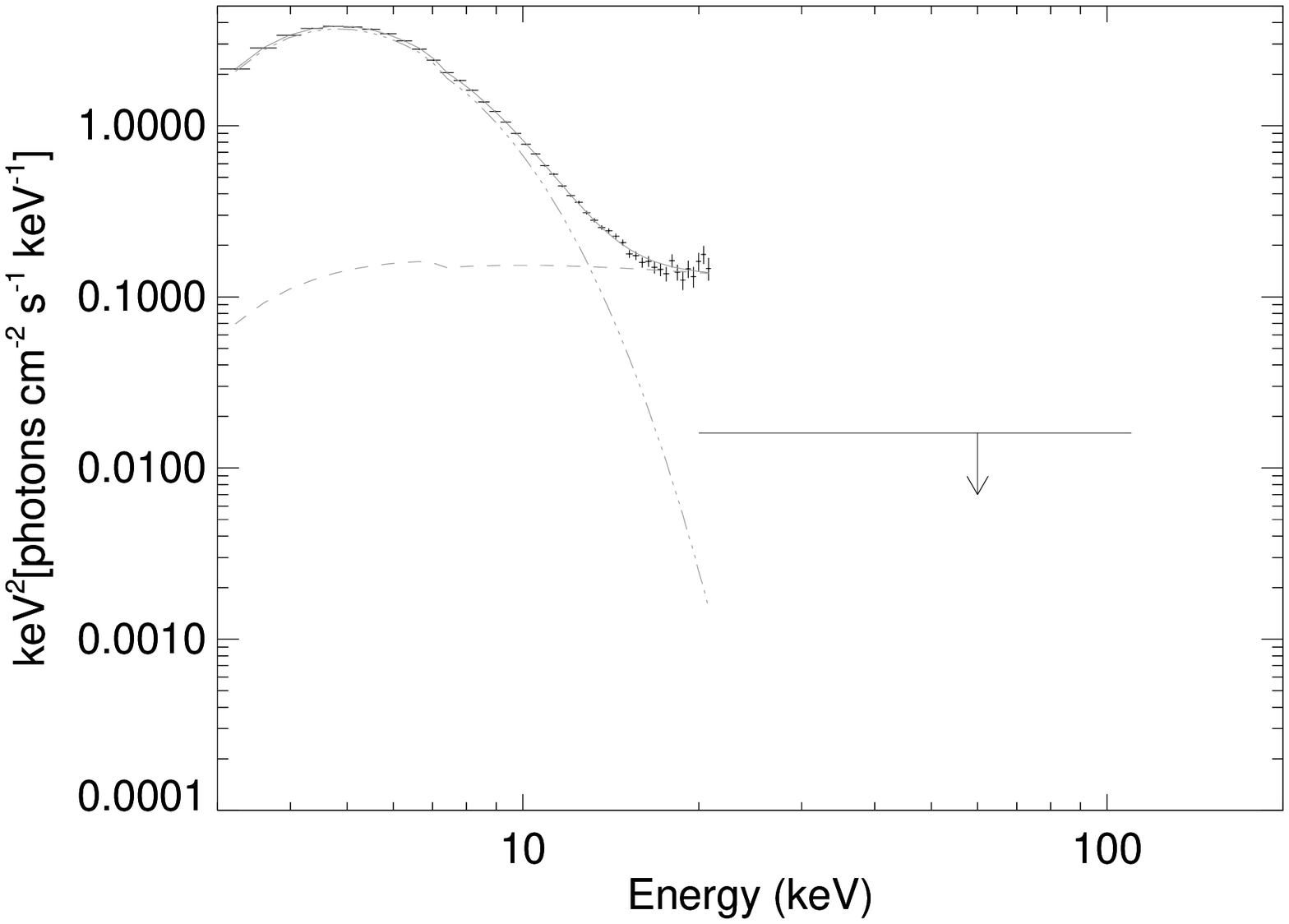}  
 \caption{{  PCA  unfolded spectrum of the 2008 outburst plus the IBIS upper limit (5mCrab)}, obs. ID 93425-01-11-07. {For model and  fitting parameters see Table~\ref{table2}. The dashed lines represent the single model components}}
\label{spettri1} 
\end{minipage}
 \ \hspace{8mm} 
 \begin{minipage}[b]{8cm}
 \includegraphics[width=9cm,angle=0]{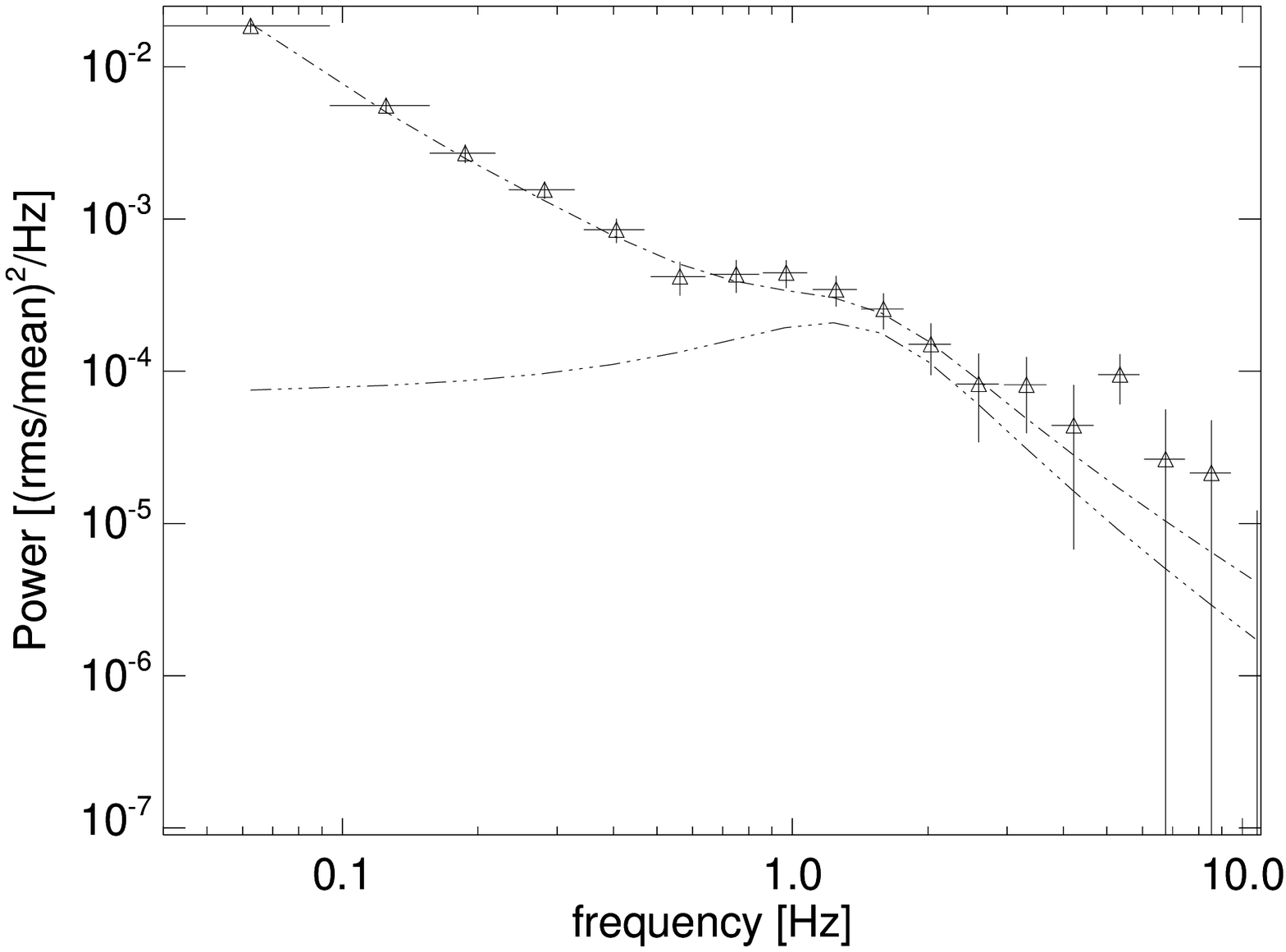}
\caption{PCA power spectrum of the 2008 outburst. {The  {\it rms} value is 0.091 (see section~\ref{se2} for details)}, obs. ID 93425-01-11-07. {For model and fitting parameters  see Table~\ref{table3}. The dashed lines represent the single model components } }
\label{spettri11}   
\end{minipage}
\begin{minipage}[b]{8cm}
\includegraphics[width=9cm,angle=0]{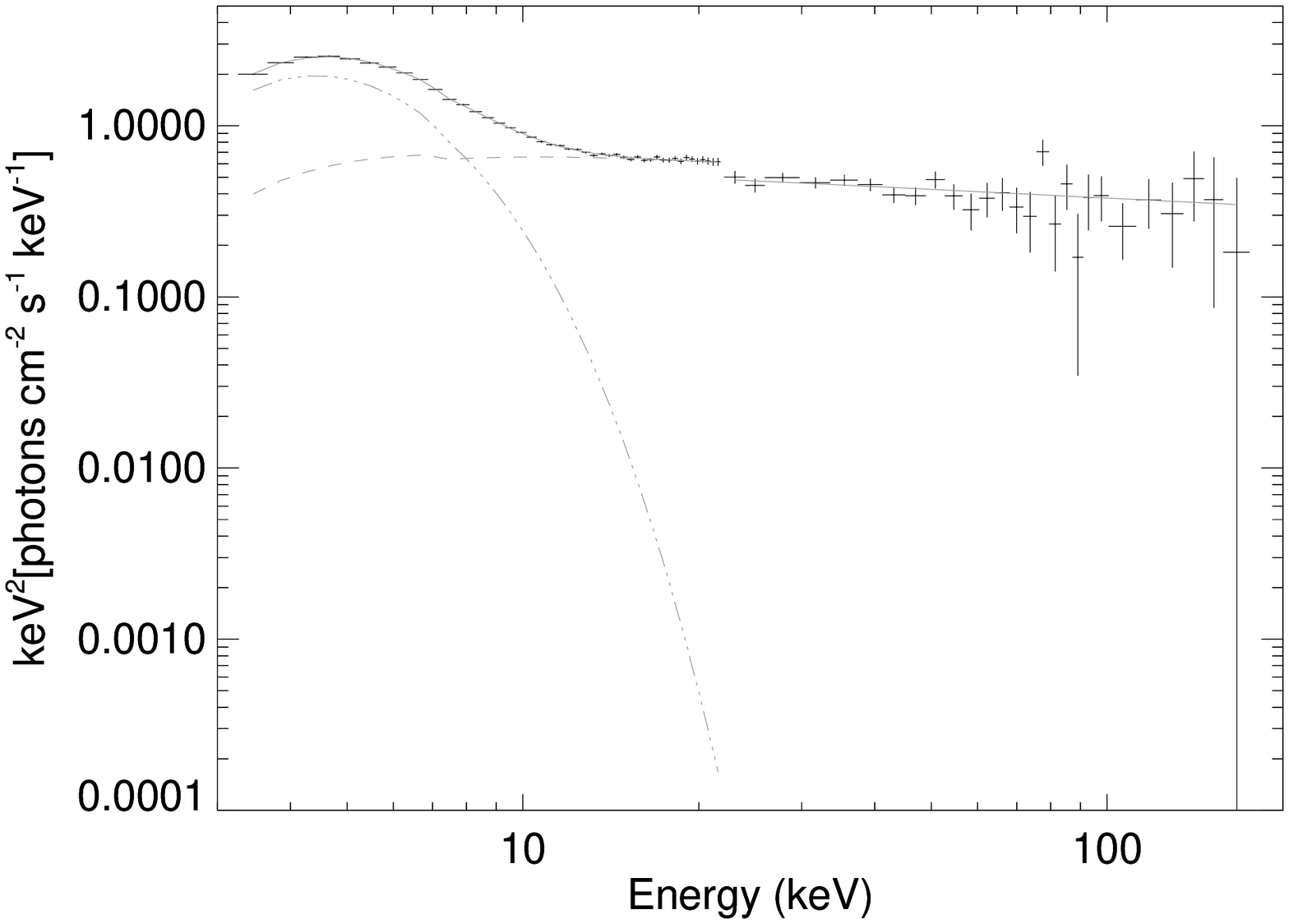}
 \caption{PCA and IBIS unfolded joint spectrum of the 2010 outburst, obs. ID: 95360-09-24-00. {for model and fitting parameters see Table~\ref{table2}. The dashed lines represent the single model components}}
\label{spettri2}  
 \end{minipage}
 \ \hspace{8mm} 
 \begin{minipage}[b]{8cm}
    \includegraphics[width=9cm]{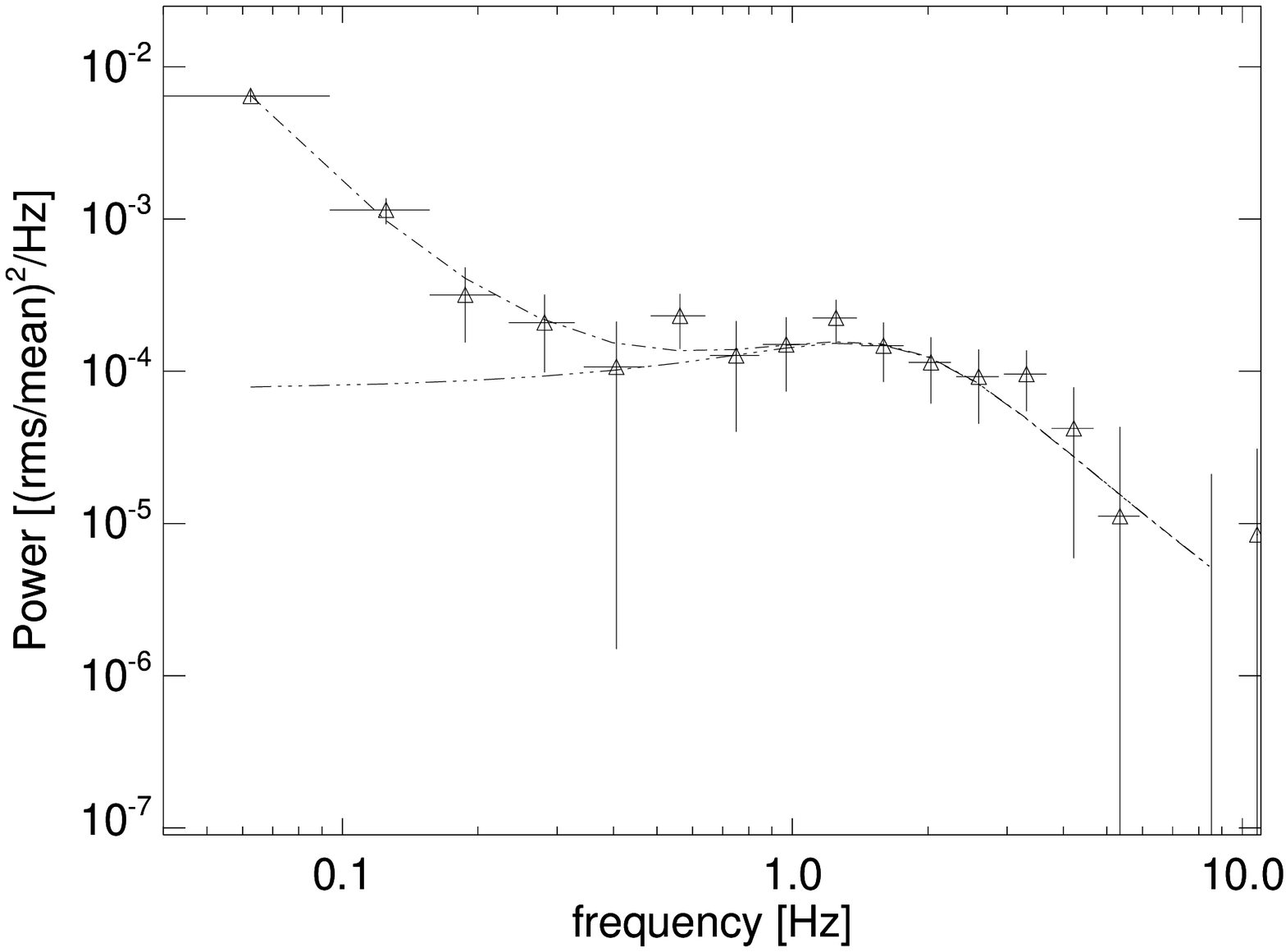}
\caption{PCA power spectrum of the 2010 outburst. {The {\it rms} value is 0.067(see section~\ref{se2} for details)}, obs. ID: 95360-09-24-00. {for model and fitting parameters see Table~\ref{table3}. The dashed lines represent the single model components}}
\label{spettri22}
 \end{minipage}
\end{figure*}

\begin{table*}
\begin{center}
\renewcommand{\thefootnote}{\alph{footnote}}

\caption{ Fit parameters of the  two energy spectra plotted in Figures~\ref{spettri1},~\ref{spettri2}. Best fit model: {\sc const*wabs*(diskbb+pow)}. {\it PCA obs. id}  is the PCA observation number; T$_{in}$ is the temperature of the inner disk; {N$_{disk}$=$((R_{in}/Km)/(D/kpc))^{2}$} is the normalization parameter of the Sakura\&Sunyaev disk black body; $\Gamma$ is the power law photon index; N$_{pow}$ is the normalization parameter of the power law model. The hydrogen column density is let free to vary between 4--12$\times$10$^{22}$ atm~cm$^{-2}$ as previously reported by~\citet{Tomsick})}
\label{table2}
\begin{tabular}{lccccccc}
\hline
\hline
PCA obs. ID  &  T$_{in}$         &       N$_{disk}$       &  $\Gamma$            &    N$_{pow}$           & Flux$_{2-10keV}$  & Flux$_{10-100keV}$ & $\chi^{2}_{red}$(d.o.f.) \\
--               &    keV               &        --                         &      --                       &    ph keV$^{-1}$ cm$^{-2}$ s$^{-1}$      &ergs~cm$^{-2}$~s$^{-1}$ &  ergs~cm$^{-2}$~s$^{-1}$& -- \\

\hline
  93425-01-11-07 & 1.31$\pm$0.01 & 340$\pm$13 & 2.3$\pm$0.2 & 0.3$^{+0.2}_{-0.1}$& 1.3$\times$10$^{-8}$ & --  & 1.1(52)    \\
  95360-09-24-00    & 1.21$\pm$0.02 &243$^{+20}_{-18}$ &2.2$\pm$0.1 & 1.0$\pm$0.2&0.8$\times$10$^{-8}$ & 1.9$\times$10$^{-9}$ &  0.8(66) \\

\hline
\hline

\end{tabular}
\caption{  Fit parameters of the  two power spectra plotted in Figures~\ref{spettri11},~\ref{spettri22}. Best fit model: (power law+Lorentzian). {\it PCA obs. ID}  is the PCA observation number;  $\Gamma$ is the power law photon index; E$_{L}$ is the Lorentzian line energy; $\sigma$ is the Lorentzian FWHM line width}
\label{table3}
\begin{tabular}{lcccccccc}
\hline
\hline
PCA obs. ID  & $\Gamma$ & E$_{L}$ & $\sigma$ & {rms}   &$\chi^{2}_{red}$(d.o.f.)\\
-- &  --& keV &keV&-- &  -- \\
\hline†
93425-01-11-07 & 1.7$\pm$0.2 & 1.2$^{+0.4}_{-1.2}$ & 1.7$^{+4}_{-1}$& 0.091 & 0.8(13)\\
95360-09-24-00 & 2.1$^{+0.5}_{-0.3}$& 1.4$^{+0.6}_{-1}$& 2.6$^{+2}_{-1}$& 0.067 & 0.5(12)\\

\hline
\hline
\end{tabular}

\end{center}

\end{table*}

\section{Discussion}
\label{se7}
\subsection{2006 and 2008 outbursts}
\label{se71}
{ The 2006 and 2008 outbursts are very similar to each other with respect to the flux emission, the outburst duration and the spectral evolution.} 
Although the PCA observation campaign began immediately after the RXTE/ASM flux enhancement, there is no evidence in the data  {of both outbursts} of the initial transition from the hard state to the soft state, as {  clearly indicated} by the HID, the HR and the {\it rms}  behaviour (see Figures~\ref{HID} and~\ref{curvaluce}). {Thus, the initial transition should have occurred shortly after the beginning of each outburst} 
or has not occurred at all. In support of this hypothesis, {since the first observation of the ATCA campaign, performed at the beginning of the 2006 outburst on  2005 December 21~\citep{Gallo}},  did not detect any radio emission. {The quenching of the radio emission is typical of HSS (see for example~\cite{Fender_1,Fender_2,Russel, Gallo_1} for a review of radio properties of XRB). 
Besides, it is well known that \bhc emits in the radio band as mentioned in  Section~\ref{se0},  during the 2002--2004 and 2013 outbursts and also during the 1998 outburst when \bhc shown X-ray luminosity comparable with the 2006 and 2008 outbursts (see Figure~\ref{HR_tot}). 
  
For the 2006 outburst, the radio non-detection fixes} with good confidence that \bhc probably reached the soft state very sharply. In fact the first radio observation was performed only  7 d after the first ASM-confirmed detection~\citep{Tomsick05_1} and 11 d before the first PCA observation (2006 January 01).

Another unusual behaviour connects these two outbursts: the lack of any detection above 30 keV all over the outburst duration.
The data show a hardening of the 3--30 keV spectra during the final phase of the two outbursts, which corresponds to an increase of the {\it rms} in agreement with what expected for a BHXRB coming back to the low-hard state (LHS)~\citep{Munoz}, even though the final LHS, if occurred, is not detected in the data. In fact, the disk black body component is always {required in the model to fit the data and the power-law component, even though very faint, seems to be typical of a soft state at least until 30 keV. Above 30 keV the non--detection implies the presence of an energy cut-off in the power law that is, instead, not typical of a soft state.} However the source flux drops rapidly, thus the final hard state transition is probably under the detection threshold of IBIS and BAT {and the PCA data are contaminated by the nearby source \igr}. 

{The lack of bright hard state has also been observed previously in \bhc~\citep[see e.g.][and references therein]{Tomsick05}. This is a unexpected behaviour, in fact, although the LHS--HSS transition can occur at different luminosity levels~\citep[see e.g.][and references therein]{Maccarone}, the initial LHS--HSS transition should occur at a luminosity about a factor of few larger than the final {HSS--HS} transition~\citep{Zdziarski04}. Thus, if the initial transition is below the BAT detection sensitivity (10.6 mCrab for 1 d of observation at 2$\sigma$), the final hard state should be really sub-luminous ($<$ 0.5--0.05 mCrab) and near to the quiescent luminosity. An explanation could simply be that the initial transition was not so sub-luminous, but instead it was so sharp that the hard X-ray monitors (IBIS and BAT) had not enough exposure time to detect it, while the PCA observation campaign started too late to observe the LHS. This is a possibility that does not involve any breaking of the hysteresis-like cycle of the HID~\citep{Homan05,Zdziarski04}.}
It remains to be explained why the {high energy emission that is thought to be due to the inverse Compton} was so faint during these two outbursts.

As demonstrated in the discussion above, the PCA energy and power spectra indicate that the source should be in a soft or intermediate soft state during the observation campaign.
 Hence, {according to ~\citet{Zdziarski04}, the accretion disk} should be extended near the last stable orbit and the electron corona located in active regions on the disk surface.  In this state we expect non thermal or hybrid distributions of the corona electrons. {This implies, at first approximation, the presence in the spectral model of a power law without cut-off. Therefore, the presence of this power law component in  the 3--30 keV} spectra and the lack of any detection above 30 keV, implies, {instead, a cut-off at energies not higher than 30 keV. This could imply that the corona should have  been, for some unknown reasons, quickly thermalised or it has not heated up at all at the beginning of the outburst, leading to a very low electron temperature,  roughly $kT_{e} \le 15$ keV {  \citep{Miyakawa,Petrucci}} . 

{However the evolution of the plasma optical depth, $\tau_{p}$ is strongly related to the fixed value of the plasma temperature kT$_{e}$. In fact the only information that can be extrapolated from the data is that the kT$_{e}$ should be less than about 10--15 keV, but its variation during the two outbursts is totally unknown.}

 \subsection{2010 outburst}
 \label{se4.2}
 The scenario is different in the case of the 2010 outburst. As in the two previous outbursts, the lack of any radio detection~\citep{Calvelo} probably implies a very sharp transition to the HSS, just 5 d before the first MAXI telescope detection reported by~\cite{Tomida}. However, the source brightens above 30 keV {not during the hard state, if occurrs, but rather during the soft part of the outburst. In fact, as can be easily seen in Figure~\ref{asm_bat}, both the soft and the hard light curves (top and bottom panel, respectively) reach the peak of emission at the same time. Normally the hard peak of the X-ray light curve (i. e. 15--50 keV) precedes the soft peak (2-12 keV)}.

{Either way, we can argue that the inverse Compton emission}, during the 2010 outburst, is more efficient than in the previous two cases. The PCA-IBIS joined spectra {  do not present} any cut-off of the power law until 150--200 keV. This means that {just from the beginning of the outburst,} the electron plasma temperature is very high or the electrons have a non thermal distribution~\citep{Malzac}, in agreement with the standard behaviour of BHXRB in the soft state ~\citep[see e.g.][]{Zdziarski04}. {However the spectral resolution does not permit to distinguish the two cases}. {Nevertheless, as reported by~\citet{Tomsick14}, the 2010 outburst shows another peculiarity}: an anomalous delay of the final transition to the LHS.
 However the last IBIS data, simultaneous with the Suzaku observations reported by~\citet{Tomsick14}, do not show the typical LHS {high energy spectrum above 20 keV.} In fact the power law remains quite soft ($\Gamma\sim$2) and there is no cut-off up to 150--200 keV. It is also true that the faintness of the data force to a very long integration time and the harder data, that should be observed simultaneously with the transition, could have been averaged out with the other observations. Data obtained with smaller integration times around the date of the HSS to LHS transition are not statistically significant.

\subsection{The periodical outbursts}
\label{se_new}
 {The  recurringof an outburst every 600--700 d could be caused by a perturbation of periodical nature which could force the system to undergo an outburst.}
 
 {We argue that there is a connection between this periodicity of the outbursts and the peculiarity of their behaviour.
In fact,  the outbursts are thought to be caused by a variation of the mass accretion rate from the companion star via Roche lobe overflow.  This variation causes an increase of the disk temperature inducing the ionization instability of the disk. In the case of \bhc there should be an extra perturbation of the system that periodically triggers this variation. This perturbation forces an increase of the disk temperature independent from the mass accretion rate of the companion star and from the state of the accretion disk and the relativistic electrons inflow. This scenario could explain also the anomalies of the LHS--HSS and HSS--LHS transitions of \bhc and the lack of bright hard state. In fact, as demonstrated by \citet{Homan}, while the variation of the mass accretion rate should cause the outbursts, the corona inflow seems to be involved in the state transitions of the BHXRB. thus we could argue that the forced outbursts do not respect all the steps that are involved in the hysteresis--like behaviour of the standard outbursts of the BHXRBs.
 
 Following this scenario the 2002-2004 bright outburst (Figure~\ref{HR_tot}) could be considered a standard outburst of \bhc due to the variation of the mass accretion rate of the companion star. This outburst covers a time period for which {the source should have undergone an outburst} two times. Surprisingly, two of the bright flux peaks of this outburst lie exactly at the epochs in which we should expect two periodical outbursts (see Figure~\ref{HR_tot}). {This fact fixes with good confidence that the unknown periodic perturbation was still active during the outburst and not related to the outburst origin.} 
 
In the last 20 yr a wide variety of hypotheses have been put forward to explain the nature of the periodical perturbation. here we critically review all the hypotheses trying to figure out the most plausible:} 

\begin{itemize} 
 
 \item {A limit cycle of accretion disk ionization instability should produce periodical outbursts as in the case of the  dwarf novae~\citep[see e. g.][]{Cannizzo, Janiuk}.
 Concerning this hypothesis, the duration and the recurrence period of the \bhc outbursts seem to be plausible for a limit cycle~\citep{Lasota}. However their time profile is not always the one expected for a limit cycle, ``fast rise and exponential decay"~\citep[FRED][]{Lasota}. As reported by e. g.~\citet{Janiuk} other factors such as the accretion rate from the companion star could  superimpose other outbursts to those produced by the limit cycle in analogy with the case of dwarf novae super-outbursts (e. g. the 2002--2004 outburst).
 Nevertheless, as~\citet{Parmar} noticed, it is unlikely that the instability could maintain the phase for more than 40 years, while this phenomena seems {to be} more plausible in the case of H 1743-322~\citep[e.g.][]{Capitanio_1} which usually shows three or four outbursts with the same period and then the next few outbursts show a different period. }  

\item {Another hypothesis could be the presence of a third body orbiting around the binary system in a hierarchical configuration~\citep{Parmar}, that is  a body moving on an approximately Keplerian orbit around the barycenter of the binary system.}  Figure~\ref{kepler} {  represents} the derived
 relation, given by the Kepler laws, between the periastron distance of a body, orbiting around a binary, with a 600-d period, as a 
 function of its mass. We roughly consider the binary system as only one heavy body of 10 M$_{\odot}$  {consistent with the result reported by~\citet{Seifina}} . We perform this  calculation for different eccentricity values of the orbit, finding that, for a highly eccentric orbit (e$>$0.7), this third body could reach a 
 distance lower than an astronomical unit (au) and its mass could be even very high. {However the three-body orbital stability criteria~\citep[see][and reference therein]{Donnison} imply that the binary remains stable against the perturbation of the third body if it is at a distance larger than a critical distance~\citep{Harrington, Donnison}. This distance strongly depends on the separation between the components of the binary system.
 As reported by~\citet{Augusteijn}, the infrared properties of \bhc are similar to those of other BHC such as
GROJ1655--40 or SAXJ1819.3--2525 that host a relatively early type secondary companion.
 {Thus the distance between the companion star and the black hole could be roughly 0.1 AU~\citep[see e. g.][]{Remillard}. {  The dashed and dotted line} in Figure~\ref{kepler} shows the critical periastron distance in function of the third body mass for a binary system separation of 0.1 AU and a prograde motion of the third body following the equation {\it (1)} of~\citet{Donnison} and the results reported by~\citet{Harrington}~\footnote[2]{ The critical distance in the case of retrograde motion is quite similar to the prograde one and is omitted for clarity }. 
 This calculation strongly depends on the binary separation which for \bhc is substantially unknown.  
 For example the black dotted line in Figure~\ref{kepler} represents the critical distance as a function of the third body mass in the case of a binary separation of 0.5 U.A. 
 A  hierarchical eccentric  third-body system  near the instability limit could be a good candidate to cause periodical outbursts. {In fact, in this case, the presence of the third body perturbs the central binary system and when it reaches the periastron, its vicinity could induce the instabilities in the accretion disk triggering an outburst. 
 Because there are too many unknown parameters, it is difficult to determine if  the sporadic increase of the recurrence period is compatible with an orbit perturbation of the triple system.  {Nevertheless it is possible to hypothesize} that the perturbation of the orbital period could be due to an effect  such us the Kozai effect, a periodic exchange between the orbit  inclination and the orbit eccentricity of the third body}~\citep[][]{Kozai}.  Thus the diagram in Figure~\ref{kepler} should be considered as a first approximation of the problem. In fact, we did not taken into account the relativistic effects and the fact that the mass of the third body could be also comparable with the other two. In this case the dynamic of a three-body system containing three comparable masses, remains a major unsolved problem~\citep{Donnison92}. However \bhc is not the first triple system containing an accreting binary. For example there are several millisecond pulsars that belong to a hierarchical 3$^{rd}$-body systems such as B1620--26, a millisecond pulsar with a white dwarf and a planetary-mass object~\citep{{Thorsett}} or the recently published PSR J0337+1715, containing a millisecond pulsars and two other stars in a hierarchical configuration~\citep{Ramson}. The presence of a third--body in this kind of systems is easy to detect and to study because of the typical time-delay of the spin period of the pulsars~\citep[see][and references therein]{Ramson}}
 
\item {We could also suppose a system with a high eccentric orbit of the companion star. However this should imply a massive companion star~\citep{Parmar} in a high mass X-ray binary (HMXRB) system. {Even though the IR could not exclude the Be nature of the system,} the spectral behaviour, the duration of the outbursts and the signature of the presence of a BH as compact object~\citep{Seifina} make this hypotheses unlike. In fact the first case of Be HMXRB hosting a BH has recently been reported in literature~\citep{Munar-Adrover}.  This source shows a spectral behavior  totally different from the behaviour of \bhcc 
\item We could consider a LMXRB in a wide eccentric orbit, which is unacceptable on a evolutionary point of view. In fact the LMXBs are old systems with circular orbits of periods that spans from a few hours (ultra-compact sources) to some days.  Due to the huge orbital period (600--700 d), even a subsequent capture of the companion star or a modification of its orbit due to interactions with other bodies  are not consistent with the Roche lobe overflow phenomena. In fact the companion star of a binary system such as  \bhc should be very small~\citep{Augusteijn} and needs to be very near to the central BH to allow the accretion coming from the Lagrange point~\citep[][d$<$0.1 AU]{Remillard}. Thus, considering the Keplerian orbits in Figure~\ref{kepler}  as the possible orbits of an object captured by a 10M$_{\odot}$ BH, such a small periastron distance could be reached only with a degenerate eccentricity of the orbit.

\item Also the constant refilling from the companion star  as reported by ~\citet{Trudolyubov} is quite unlikely. In fact,  even accepting that a constant refilling of the outer disk could cause the periodical outbursts, this implies a constant quantity of accreted matter that in some way reaches, after about 600 days, a critical mass that, in turn, triggers the outburst, involving always the same quantity of matter. Thus the outbursts should have always the same luminosity. As Figure~\ref{HR_tot} shows this is not the case of \bhcc }}
\end{itemize}

\begin{figure}
\includegraphics[scale=0.43,angle=0]{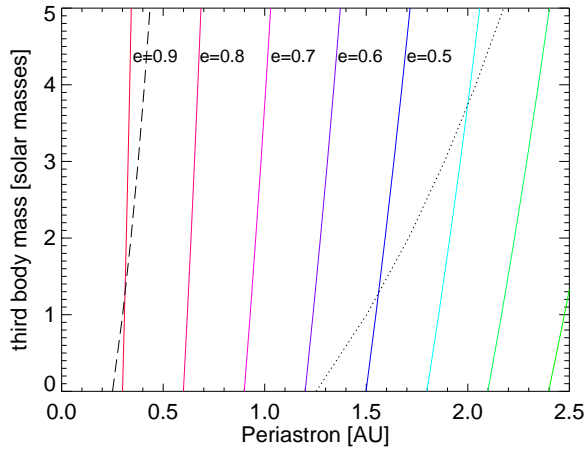}
\caption{Periastron distance of a body orbiting around the binary star (period: 600 d) in function of its mass. The plot reports this relation at different values of eccentricity. We assumed the XRB as a only one heavy body of M=10M$_{\odot}$.  Considering  a prograde motion of the third body: the black dashed line represents the critical distance for a binary separation of 0.1 AU and the black dotted line represents the critical distance for a binary separation of 0.5 AU. The  calculation of the critical distances has been performed following the equation {\it (1)} of ~\citet{Donnison} ~\citep[see also][for details]{Harrington}.}
\label{kepler}
\end{figure}

\section{Conclusions}
\label{se8}

\begin{enumerate}
\item  {The  2006 and 2008 outbursts exhibit an unusual behaviour: they do not show any hard emission up to 30 keV throughout the outburst. Probably the electron corona is quickly thermalized or it has not enough time to heat up.  Moreover at least in the 2006 case, the lack of any radio detection immediately after the enhancement of the ASM flux indicates that the LHS--HSS transition occurred very sharply  or did not occur at all.  This is quite a rare behaviour for a LMXB. In fact, outbursts that never reach a soft state (the so called "failed outbursts") have often been reported in literature~\citep[see e.g.][and references there in]{Capitanio09,Ferrigno}. }
{ On the contrary "soft" outbursts are  unusual since the launch of the X-ray satellite hosting telescopes suitable for hard X-ray emission measurement above 20 keV, such as BeppoSAX/PDS, RXTE/HEXTE, INTEGRAL/IBIS and Swift/BAT}

\item The 2010 outburst seems, in contrast, to be a standard outburst from the high energy emission point of view. 
In fact the source is bright above 30 keV for most of the outburst, although, in this case also, as reported by~\citet{Tomsick14}, the outburst shows an unusual delay of the final transition to the hard state. {Furthermore, the lack of radio emission in the rising phase of the outburst indicates a sharp LHS--HSS transition.}

\item The 2006 and 2008 outbursts of \bhc do not seem to follow the empirical relations found by ~\citet{Wu}, ~\citet{Yu} and ~\citet{Yu2} between the hard X-ray  luminosity peak of the HS--HSS transition and the subsequent HSS luminosity peak, or the outburst waiting time.  These empirical relations imply that the two  accretion flows (disk and corona) should be related. Our results give an indication that at least in these cases the two accretion flows could be uncorrelated.

 \item  {We argue that an important role in the anomalous behaviour of \bhc could be played by the periodical outbursts, whatever their origin. The periodicity could be an extra perturbation that is added to the disk instability due to the companion star mass transfer. The interference between these two phenomena could explain for example the long and bright outburst occurred between 2002 and 2004.}

\item {As we illustrated in Section~\ref{se_new}, there are tow most plausible explanations for the outbursts being equally spaced in time. The first is the presence of a third body orbiting around the binary system, while the second is the presence of a limit cycle of the accretion disk ionization instability.
Concerning the latter hypothesis, it is unlikely that this limit cycle phase could have lasted for such a long time.
Instead, considering the third body hypothesis, we have discussed the possibility that the recurrence period is compatible with a three-body hierarchical system containing an X-ray binary and a third body that could be even quite heavy and near the critical distance. {Nevertheless}, these calculations strongly depend on the binary system separation that is actually unknown.
At odd with this hypothesis is the sporadic 100-d longer outburst recurrence period, previously noticed by other authors but also present in the last three outbursts of the source. However our simple calculation does not take into account the effect of the space-time deformation on the third body due to the BH and the various classical perturbations that affects a hierarchical three body-system such as the Kozai effect. In fact the unknown nature of the binary and of the third body leave too much unknown parameters to permit any more refined calculations. }

\end{enumerate}

 We can conclude that, even if we have demonstrated that the case of \bhc could be peculiar with respect to the other BHCs,  the origin and evolution of the electron corona is a complex mechanism not only related to the spectral evolution of the accretion disk and still not well understood. 
 
\section*{ACKNOWLEDGEMENTS}
{ The authors would to thank Roberto Ricci and Marcello Giroletti for checking the radio data. RC and the other authors thank T.M Belloni for his help with timing analysis software. FC also thank Agnieszka Janiuk, Massimo Cocchi, Agata Rozanska, Melania Del Santo and Carla Maceroni for useful scientific discussions. A special thanks goes to Valeria Mangano for scientific discussions and careful editing of the text.   The RXTE data were obtained through the High Energy Astrophysics Science Archive Research Center ({\it HEASARC}) online service. The INTEGRAL data were obtained through the INTEGRAL Science Data Center ({\it ISDC}) online service. The  Swift/BAT light curve is part of  the SWIFT/BAT transient monitor results provided by the Swift/BAT team.}

\end{document}